\newif\ifOneCol

\ifOneCol
    \documentclass[12pt,onecolumn,draftcls]{IEEEtran}
    \renewcommand{\baselinestretch}{1.48}
\else
    \documentclass[journal,10pt,twocolumn]{IEEEtran}
    \renewcommand{\baselinestretch}{0.96}
\fi

\usepackage{graphicx}
\usepackage{epsfig}
\usepackage{subfigure}
\usepackage{amssymb}
\usepackage{amsbsy}
\usepackage{amsmath}
\usepackage{steinmetz}
\usepackage{algorithmic}
\usepackage{tabularx}
\usepackage{algorithm}
\usepackage[font=scriptsize]{caption}

\usepackage{cite}
\usepackage{url}

\usepackage{amsfonts}

\usepackage{color}
\usepackage{float}
\usepackage{times,amsmath,epsfig}
\usepackage{xspace,latexsym,syntonly}
\usepackage{amssymb}
\usepackage{textcomp}

\usepackage{xcolor}

\newtheorem{definition}{Definition}
\newtheorem{theorem}{Theorem}
\newtheorem{remark}{Remark}
\newtheorem{lemma}{Lemma}

\newcommand{\beq}{\begin{equation}}
\newcommand{\eeq}{\end{equation}}
\newcommand{\bea}{\begin{array}}
\newcommand{\ena}{\end{array}}
\newcommand{\bds}{\begin {itemize}}
\newcommand{\eds}{\end {itemize}}
\newcommand{\bdf}{\begin{definition}}
\newcommand{\blm}{\begin{lemma}}
\newcommand{\edf}{\end{definition}}
\newcommand{\elm}{\end{lemma}}
\newcommand{\bthm}{\begin{theorem}}
\newcommand{\ethm}{\end{theorem}}
\newcommand{\bprp}{\begin{prop}}
\newcommand{\eprp}{\end{prop}}
\newcommand{\bcl}{\begin{claim}}
\newcommand{\ecl}{\end{claim}}
\newcommand{\bcr}{\begin{coro}}
\newcommand{\ecr}{\end{coro}}
\newcommand{\bquest}{\begin{question}}
\newcommand{\equest}{\end{question}}

\newcommand{\larrow}{{\larrow}}

\newcommand{\nin}{{\not \in}}

\def\urltilda{\kern -.15em\lower .7ex\hbox{\~{}}\kern .04em}

\begin{document}\title{Distributed Learning over Markovian Fading Channels for Stable Spectrum Access}
\author{Tomer Gafni and Kobi Cohen
\thanks{Tomer Gafni and Kobi Cohen are with the School of Electrical and Computer Engineering, Ben-Gurion University of the Negev, Beer Sheva 8410501 Israel. Email: gafnito@post.bgu.ac.il, yakovsec@bgu.ac.il.}
\thanks{This work has been submitted to the IEEE for possible publication. Copyright may be transferred without notice, after which this version may
no longer be accessible.}
\thanks{A short version of this paper was presented at the 57th Annual Allerton Conference on Communication, Control, and Computing, 2019 \cite{gafni2019distributed}.}
}
\date{}
\maketitle
\begin{abstract}
\label{sec:abstract}
We consider the problem of multi-user spectrum access in wireless networks. The bandwidth is divided into $K$ orthogonal channels, and $M$ users aim to access the spectrum. Each user chooses a single channel for transmission at each time slot. The state of each channel is modeled by a restless unknown Markovian process. Previous studies have analyzed a special case of this setting, in which each channel yields the same expected rate for all users. By contrast, we consider a more general and practical model, where each channel yields a different expected rate for each user. This model adds a significant challenge of how to efficiently learn a channel allocation in a distributed manner to yield a global system-wide objective. We adopt the stable matching utility as the system objective, which is known to yield strong performance in multichannel wireless networks, and develop a novel Distributed Stable Strategy Learning (DSSL) algorithm to achieve the objective. We prove theoretically that DSSL converges to the stable matching allocation, and the regret, defined as the loss in total rate with respect to the stable matching solution, has a logarithmic order with time. Finally, simulation results demonstrate the strong performance of the DSSL algorithm.
\end{abstract}
%

\section{Introduction}
\label{sec:introduction}
We consider the spectrum access problem, where a shared bandwidth is divided into $K$ orthogonal channels (i.e., sub-bands), and $M$ users want to access the spectrum, where $K\geq M$. Each channel is modeled by a Finite-State Markovian Channel (FSMC), which is independent and non-identically distributed across channels. The FSMC is a tractable model widely used to capture the time-varying behavior of a radio communication channel \cite{wang1995finite,sadeghi2008finite}. It is often employed to model radio channel dynamics due to primary user occupancy effects in hierarchical cognitive radio networks (where the $M$ secondary (unlicensed) users are cognitive in terms of learning and adapting good access strategies), or the external interference effects in the open sharing model among $M$ users in the wireless network (e.g., ISM band) \cite{zhao2007survey,slamnik2020sharing}. At each time step, each user experiences a different transmission rate over each channel depending on its FSMC distribution, where the FSMC parameters (i.e., the transition probabilities that govern the Markov chain) are unknown. At each time step, each user is allowed to choose one channel to access, and observe the instantaneous channel state. If two users or more access the same channel at the same time, a collision occurs and the achievable rate is zero.

We adopt the stable matching utility (see Section \ref{sec:problem} for details) as the system objective, which is known to yield strong performance in multichannel wireless networks \cite{leshem2012multichannel}. We define the \textit{regret} as the loss in total rate with respect to the stable matching solution with known FSMCs. The objective is to develop a distributed learning algorithm for channel allocation and access under unknown FSMCs that minimizes the growth rate of the regret with time $t$.

\subsection{Main Results}
\label{ssec:results}

The stable matching problem for multi-user spectrum access was first introduced in \cite{leshem2012multichannel} under the assumption that the expected rates are known, and a distributed opportunistic CSMA algorithm that solves the problem was proposed. The model with an unknown expected rate matrix and rested setting (i.e., the states of the Markovian process do not change if not observed by the user) was studied in \cite{kalathil2014decentralized, nayyar2016regret}. A regret (with respect to the optimal allocation) of near-$O(\log t)$ was achieved. However, these algorithms require intensive communication between users in order to apply the auction algorithm \cite{bertsekas1988auction}. In \cite{avner2016multi}, the authors reduced the communication burden, but without guarantees on the achievable regret. Recently, it was shown in \cite{bistritz2018distributed,2019arXiv190201239B} that achieving a sum-regret of near- $O(\log t)$ is possible without communication between users, but only for the case of i.i.d channels. In this paper we focus on the general case where the channel states may change whether or not they are being observed (i.e., the restless Markovian setting), and improve the regret scaling with the system parameters by a simple distributed implementation. The main contributions are summarized below.
\paragraph{A general model for spectrum access using a restless Markovian channel model} As explained above, by contrast to \cite{leshem2012multichannel, kalathil2014decentralized, nayyar2016regret, avner2016multi, bistritz2018distributed,2019arXiv190201239B}, in this paper we first solve the channel allocation and access problem under general unknown restless Markovian channel model. Handling this model adds significant challenges in algorithm design and regret analysis. Due to the restless nature of the channels and potential reward loss due to transient effects as compared to steady state when switching channels, learning the Markovian channel characteristics requires that the channels be accessed in a judicious consecutive manner for a period of time. This is reflected in a novel algorithm design that guarantees efficient learning, as detailed next.
\paragraph{Algorithm Development}
We are facing an online learning problem constituted by the well-known exploration versus exploitation dilemma. To remedy this, we propose a novel Distributed Stable Strategy Learning (DSSL) algorithm for solving the problem. Since the FSMCs are unknown, the rate means must be learned by accessing all channels via exploration phases. This results in increasing the regret, since the stable allocation is not performed. Thus, the exploration time must be minimized, while guaranteeing efficient learning. Roughly speaking, each channel can be learned by different exploration times, depending on its unknown parameters (see more details in Section \ref{ssec:parameter}). The algorithm design in this paper contributes to both tackling the more general model, as well as improving the learning efficiency in a fully-distributed manner. Specifically, in existing algorithms \cite{kalathil2014decentralized, nayyar2016regret,avner2016multi, bistritz2018distributed,2019arXiv190201239B}, the exploration phase of all channels is determined by the channel that requires the largest exploration time. This results in oversampling the channels and significantly increases the regret. By contrast, the DSSL algorithm estimates online the desired (unknown) exploration rate of each channel. Thus, by sampling the channels according to the desired exploration rate, it avoids oversampling the channels, and thus reduces the regret scaling significantly as compared to existing algorithms.
\paragraph{Performance analysis}
In terms of theoretical performance analysis, we prove that the DSSL algorithm converges to the stable matching allocation, and the regret has a logarithmic order with time. When comparing to existing approaches, DSSL achieves this under the more general restless Markovian model, and also has significantly better scaling with the system parameters. Specifically, under a common benchmark setting of equal rates among users (but still vary among channels), and $K>M$, which allows a theoretical comparison of learning efficiency between different algorithms, in \cite{nayyar2016regret} and \cite{Liu_2013_Learning} the regret scales as $O(\frac{MK}{(\Delta_{\min})^2}\log(t))$ ,in \cite{2019arXiv190201239B} as $O(\frac{M^3K}{(\Delta_{\min})^2}\log(t))$ and in \cite{bistritz2018distributed} the regret scales as $O(\frac{MK^2}{(\Delta_{\min})^2}\log(t))$, where $\Delta_{\min}$ is the difference in rates between the $M$th and $(M+1)$th best channels. In contrast, under DSSL, the regret scales as $O((\frac{1}{(\Delta_{\min})^2}+MK)\log(t))$. In addition, extensive numerical experiments were performed to demonstrate the efficiency of the proposed DSSL algorithm. 

\subsection{Related Work}
\label{ssec:related}

A number of studies have developed distributed learning algorithms for a special case of the restless Markovian channel model considered in this paper, where each channel yields the same expected rate for all users \cite{Tekin_2012_Online, liu2012learning, gafni2018learningISIT}. This special case significantly simplifies the channel allocation problem and the analysis (for instance, switching between assigned users does not affect the resulting regret in this special case). In this paper, we consider the general model where each channel yields a different expected rate for each user. This models the situation of different channel fading states across users and channels in actual wireless networks, and adds a significant challenge of how to learn the desired channel allocation in a distributed manner to achieve a global system-wide objective.

Another set of related work on multi-user channel allocation has approached it from the angle of game theoretic and congestion control (~\cite{han2005fair, menache2008rate, candogan2009competitive, menache2011network, law2012price, cohen2013game, wu2013fasa, singh2016combined, cohen2016distributedToN, cohen2017distributed, malachi2020queue} and references therein), hidden channel states\cite{yemini2019restless}, and graph coloring (\cite{wang2005list, wang2009improved,  checco2013learning, checco2014fast} and references therein).
The game theoretic aspects of the problem have been investigated from both non-cooperative (i.e., each user aims at maximizing an individual utility) \cite{menache2008rate, candogan2009competitive, singh2016combined, cohen2016distributedToN, cao2018distributed}, and cooperative (i.e., each user aims at maximizing a system-wide global utility) \cite{han2005fair, leshem2006bargaining, cohen2017distributed, bistritz2018approximate} settings. Model-free learning strategies were developed in \cite{naparstek2017deep, naparstek2018deep} for orthogonal channels, compact models \cite{livne2020pops}, and multiple access channel strategies were developed in 
\cite{sery2019analog, cohen2019time}. Graph coloring formulations have dealt with modeling the spectrum access problem as a graph coloring problem, in which users and channels are represented by vertices and colors, respectively (see \cite{wang2005list, wang2009improved, checco2014fast, checco2013learning} and references therein for related studies). Finally, none of these studies have considered the problem of achieving provable stable strategies in the learning context under unknown restless Markovian dynamics, as considered in this paper.

\section{System Model and Problem Formulation}
\label{sec:problem}

We consider a wireless network consisting of $K$ orthogonal channels indexed by the set $\mathcal{K} = \{1,2,...,K\}$ and $M$ cognitive users (referred to as users) indexed by the set $\mathcal{M} = \{1,2,...,M\}$, where $K \geq M$. The users aim at accessing the spectrum to send their data. Each user is allowed to choose a single channel for transmission at each time slot, and transmit if the channel is not occupied by a primary user. The users operate in a synchronous time-slotted fashion. Due to spatial geographic dispersion, each user can potentially experience different achievable rates over the channels. When a user $i$ transmits on channel $k$ (when the channel is free) at time slot $t$, its data rate is given by $r_{i,k}(t)$. This information is concisely represented by an $M \times K $ rate matrix $V(t) = \{r_{ik}(t)\}$, $i=1, ..., M, k=1, ..., K$.

We consider the case where the rate process $r_{i,k}(t)$ is Markovian and has a well-defined steady state distribution. The transition probabilities associated with the Markov chain are unknown to the users. The process $r_{i,k}(t)$ evolves independently of the user's actions (i.e., external process). Furthermore, the channel states may change depending on whether or not they are observed (i.e., restless setting). Specifically, the rate of user $i$ on channel $k$, $r_{i,k}(t)$, is modeled as a discrete time, irreducible and aperiodic Markov chain on a finite-state space $\mathcal{X}^{i,k}$ and is represented by a transition probability matrix $P^{i,k} \triangleq (p^{i,k}_{x,x'}: x,x' \in \mathcal{X}^{i,k})$.
The process mean (i.e., the expected rate) is denoted by $\mu_{i,k}$ and is unknown to the users. We define the $M \times K $ expected rate matrix by $U = \{\mu_{ik}\}$, $i=1, ..., M, k=1, ..., K$.

Let $X_{i,k}(t)$ be the actual achievable rate for user $i$ on channel $k$ at time $t$. If two or more users choose to access the same channel at the same time slot, a collision occurs. In this case, $X_{i,k}(t)=0$. Otherwise, if user $i$ has accessed channel $k$ without colliding with other users, then $X_{i,k}(t)=r_{i,k}(t)$. The users implement carrier sensing to observe the current channel state at each time slot as is typically done in cognitive radio networks \cite{Tekin_2012_Online, cohen2013game}. Hence, the channel states are observed regardless of collisions. The transmission scheme for the multi-user spectrum access model is detailed in Section \ref{sec:DMASR}. 

\subsection{Notations}

We present the other notations that are used throughout the paper. Let $\vec{\pi}_{i,k} \triangleq (\pi^x_{i,k}, x \in \mathcal{X}^{i,k})$ be the stationary distribution of the Markov chain $P^{i,k}$, and let:
\begin{center}
$\displaystyle\pi_{\min}\triangleq \min_{i \in \mathcal{M}, k \in \mathcal{K},x \in \mathcal{X}^{i,k}}\ \pi^x_{i,k}, \quad
\hat{\pi}_{i,k}^x \triangleq \max\{\pi_{i,k}^x,1-\pi_{i,k}^x\}, \quad \hat{\pi}_{\max} \triangleq \max_{i \in \mathcal{M}, k \in \mathcal{K},x \in \mathcal{X}^{i,k}}\{\pi_{i,k}^x,1-\pi_{i,k}^x\} $.
\end{center}
We define $X_{\max} \triangleq \max_{i \in \mathcal{M}, k \in \mathcal{K}}\{ |\mathcal{X}^{i,k}| \}$ as the maximal cardinality among the state spaces, and 
\begin{center}
$\displaystyle x_{\max}\triangleq \max_{i \in \mathcal{M}, k \in \mathcal{K},x \in \mathcal{X}^{i,k} } x, \quad r_{\max}\triangleq\max_{i \in \mathcal{M}, k \in \mathcal{K}}\sum_{x\in{\mathcal{X}^{i,k}}}x $.
\end{center}   
Let $\lambda_{i,k}$ be the second largest eigenvalue of $P^{i,k}$, and
$\displaystyle\lambda_{\max}\triangleq\max_{i \in \mathcal{M}, k \in \mathcal{K}}\ \lambda_{i,k}$
be the maximal one among all channels and users. Also,
$\displaystyle\overline{\lambda}_{\min}\triangleq 1-\lambda_{\max}, \displaystyle\overline{\lambda}_{i,k}\triangleq 1-\lambda_{i,k}$
is the eigenvalue gap.
Let $M^{i,k}_{x,y}$  be the mean hitting time of state $y$ starting at initial state $x$ for channel $k$ used by user $i$, and
$\displaystyle M^{i,k}_{\max}\triangleq\max_{x,y \in \mathcal{X}^{i,k}, x\neq y}M^{i,k}_{x,y}$.
%
We also define:
\begin{center}
$ A_{\max}\triangleq\max_{i,k}\;(\min_{ x \in \mathcal{X}^{i,k}}\ \pi_{i,k}^x)^{-1} \sum\limits_{x\in{\mathcal{X}^{i,k}}}x,$
\end{center}
and
\beq
\label{eq:L}
\bea{l}
\displaystyle L\triangleq\frac{28 x_{\max}^2 r_{\max}^2 \hat{\pi}_{\max}^2} {\bar{\lambda}_{\min}}.
\ena
\eeq
The expectations $\mu_{i,k}$ are given by: 
\begin{center}
$\displaystyle\mu_{i,k}=\sum\limits_{x\in{\mathcal{X}^{i,k}}}x \cdot \pi_{i,k}^x$,
\end{center}
and we define $\sigma_i$, for $i=1, ..., M$, as a permutation of $\{1,\ldots,K\}$ such that
\begin{center}
$\displaystyle \mu_{i,\sigma_i(1)}>\mu_{i,\sigma_i(2)}> \ldots >\mu_{i,\sigma_i(K)}$.
\end{center}

\subsection{A Stable Channel Allocation}

Let $a_i(t) \in \mathcal{K}$ be a selection rule, indicating which channel is chosen by user $i$ at time $t$, which is a mapping
from the observed history of the process (i.e., all past actions and observations up to time $t-1$) to $\left\{1, ..., K\right\}$. The expected aggregated data rate for all users up to time $t$ is given by:
\beq
\label{eq:reward}
\bea{l}
\displaystyle R(t) = \mathbb{E} [\sum\limits_{n=1}^t \sum\limits_{i=1}^M X_{i,a_i(n)}(n)].
\ena
\eeq 
A policy $\phi_i$ is a time series vector of selection rules: $\phi_i = (a_i(t), t=1, 2, ...)$ for user $i$.

\noindent
\textbf{Definition 1 (\cite{leshem2012multichannel}):} A bipartite matching between channels and users is a permutation $P: \mathcal{M} \rightarrow \mathcal{K}$. The optimal centralized allocation problem is to find a bipartite matching:
\begin{center}
$\displaystyle \mathbf{k}^{**} = \arg\max_{\mathbf{k} \in P} \sum \limits_{i=1}^M \mu_{i,k(i)}$.
\end{center}

\noindent
\textbf{Definition 2 (\cite{leshem2012multichannel}):} A matching $S:\mathcal{M} \rightarrow \mathcal{K}$ is stable if for every $i \in \mathcal{M}$ and $k \in \mathcal{K}$ satisfying $S(i) \neq k$, if $\mu_{i,S(i)} < \mu_{i,k}$ then there exists some user $i' \in \mathcal{M}$ such that $S(i')=k$ and $\mu_{i',k} > \mu_{i,k}$. 

Achieving the optimal allocation in Definition 1 requires implementing a centralized solution, or a distributed solution with heavy complexity and slow convergence \cite{naparstek2013fully}. Therefore, we are interested in developing a distributed algorithm with low complexity that converges to the stable matching solution in Definition 2 which is known to yield strong performance and very fast convergence (when the expected rates are known) by using distributed opportunistic CSMA (see Section \ref{ssec:Allocation} and \cite{leshem2012multichannel} for more details on opportunistic CSMA for stable channel allocation).

We assume that the entries in the matrix $U$ are all different, as in \cite{leshem2012multichannel}, which holds in wireless networks due to continuous-valued Shannon rates\footnote{Otherwise, we can add noise to the matrix.}. Thus, there is a unique stable matching solution under our assumptions, and the expected aggregated rate under the stable matching solution $S$ is given by: $\sum \limits_{i=1}^M \mu_{i,S(i)}$. The channel $S(i)$ (i.e., the channel that user $i$ selects under the stable matching configuration) is referred to as the \textit{stable channel selection} of user $i$. 

\begin{remark}
We point out that under an i.i.d. or rested\footnote{In the rested model the Markov chain $P^{i,k}$ makes a state transition only when user $i$ accesses channel $k$.} Markovian channel model, the optimal policy is to transmit on the same channels that achieves the optimal centralized allocation in terms of the sum expected rate. However, the optimal policy in the restless Markovian setting has been shown to be P-SPACE hard even under known Markoivan dynamics \cite{papadimitriou1999complexity}. Therefore, a commonly adopted approach in this setting is to use a weaker definition of the regret, first introduced in \cite{auer2002nonstochastic} and used later; e.g., in \cite{Tekin_2012_Online,liu2012learning,lesage2017multi,reverdy2016satisficing}, where the policy is compared to a "partially informed" genie who knows the expected rates of the channels, instead of the complete system dynamics. In this paper we adopt this approach as well. 
\end{remark}

\subsection{The Objective}
\label{ssec:objective}

Since the expected rates $\mu_{i,k} $ are unknown in our setting, the users must learn this information online effectively so as to converge to the stable matching solution. A widely used performance measure of online learning algorithms is the regret, which is defined as the reward loss with respect to an algorithm with a side information on the model. In our setting, we define the regret for policy $\phi = (\phi_i, 1 \leq i \leq M)$ as the loss in the expected aggregated data rate with respect to the stable matching solution that uses the true expected rates:
\beq
\label{eq:regret}
\bea{l}
\displaystyle r_\phi(t) \triangleq t \cdot \sum \limits_{i=1}^M \mu_{i,S(i)} - \mathbb{E}_\phi [\sum\limits_{n=1}^t \sum\limits_{i=1}^M X_{i,\phi_i(n)}(n) ].
\ena
\eeq 
A policy $\phi$ that achieves a sublinear scaling rate of the regret with time (and consequently the time averaged regret tends to zero) approaches the required stable matching solution. The essence of the problem is thus to design an algorithm that learns the unknown expected rates efficiently to achieve the best sublinear scaling of the regret with time.

\section{The Distributed Stable Strategy Learning (DSSL) Algorithm}
\label{sec:DMASR}

To achieve the objective, as detailed in Section \ref{ssec:objective}, we divide the time horizon into three phases, we term exploration, allocation, and exploitation. These three phases are performed repeatedly during the algorithm according to judiciously designed policy rules, as detailed later. 

The purpose of the exploration phase is to allow each user to explore all the channels to identify its $M$ best channels (i.e., the $M$ channels that yield the highest expected rates for the user). The users use the sample means as estimators for the expected rates of the channels to achieve this goal. This phase results in a regret loss, since users access sub-optimal channels to explore them, and the stable allocation is not performed. However, this phase is essential to identifying the $M$ best channels and consequently minimizing the regret scaling with time. The purpose of the exploitation phase is to use the currently learned information to execute the stable matching solution. The allocation phase allows users to allocate the channels among themselves properly in a distributed manner using opportunistic carrier sensing \cite{zhao2005opportunistic}.

Since the rate process $r_{i,k}(t)$ can evolve even when channel $k$ is not selected by user $i$, learning the Markovian rate statistics requires using the channels in a consecutive manner for a period of time \cite{Tekin_2012_Online, liu2012learning}. Moreover, frequent switching between channels can cause a loss due to the transient effect. The high-level structure of the DSSL algorithm works as follows. Each user $i$ computes its sufficient number of samples in the exploration phases (condition (\ref{eq:Condition}) defined in \ref{ssec:selection}) for each channel $k$ at the end of every exploitation phase $t$. If the number of samples is greater than the required number for all $k$, user $i$ performs another exploitation phase. Otherwise, if the number of samples is smaller than the sufficient number for one or more channels, user $i$ carries out an exploration phase for those channels. When no exploration phase is needed, an allocation phase is performed. At the end of the allocation phase, each user identifies its stable channel selection, and an exploitation phase is carried out. We now discuss the structure of the DSSL algorithm in details.

\subsection{The structure of the exploration phase:}
\label{ssec:exploration}

Let $n_O^{i,k}(t)$ be the number of exploration phases in which channel $k$ was selected by user $i$ up to time $t$. Each exploration phase is divided into two sub epochs: a Random size Epoch (RE), and a Deterministic size Epoch (DE). Let $\gamma^{i,k}(n_O^{i,k}(t)-1)$ be the last channel state observed at the $(n_O^{i,k}(t)-1)^{th}$ exploration phase. RE starts at the beginning of the exploration phase until state $\gamma^{i,k}(n_O^{i,k}(t)-1)$ is observed. This epoch ensures that the generated sample path (after removing the samples observed in the RE epochs) is equivalent to a sample path generated by continuously sensing the Markovian channel without switching. This step guarantees a consistent estimation of the expected rates. Then, DE starts by sensing the channel for a deterministic period of time $4^{n_O^{i,k}(t)}$. The deterministic period of time grows geometrically with time to ensure a relatively small number of channel switching. 
\subsection{The structure of the allocation phase:}
\label{ssec:Allocation}

The allocation phase applies opportunistic CSMA among users. In opportunistic CSMA, the backoff function maps from an index (i.e., expected rate) to a backoff time \cite{zhao2005opportunistic}. The backoff function decreases monotonically with the rates, so that the user with the highest rate on a certain channel waits the minimal time before transmission. All other users sense that the channel is occupied and do not transmit on that channel. To obtain the stable matching allocation, this procedure continues until all $M$ users occupy $M$ channels. For more details on opportunistic CSMA for stable matching see \cite{leshem2012multichannel}. 

The allocation phase has two goals in our setting. The first is to assign channels to users to yield a stable matching solution as in \cite{leshem2012multichannel}. However, since the expected rates are unknown in our setting, the allocation phase is executed by using the sample means. 
The second goal is to use the backoff function to identify the differences in sample means among users and channels, which is needed for setting efficient learning rates. This requires a new mechanism that  performs opportunistic CSMA, as detailed below. 

Let $\mathcal{T}_k$ be the set of all users that attempt to transmit on channel $k$ at a certain stage of the allocation phase. We initialize the phase by declaring each user to be \textit{unassigned}. We divide the time horizon of the allocation phase into two sub-phases. In the first sub-phase, referred to as $S_1$, we perform opportunistic CSMA for stable matching as in \cite{leshem2012multichannel}, while replacing the expected rates by the sample means. Specifically, each unassigned user attempts to transmit on its best channel, out of those it has not yet attempted using opportunistic CSMA. On each channel $k$, the best user out of $\mathcal{T}_k$ in this sub-phase ($S_1$) is declared to be assigned. All the other users in $\mathcal{T}_k$ store the sample mean of the assigned user (by mapping from the sensed backoff time to the sample mean). This sub-phase continues until all $M$ users are assigned to $M$ channels.
The second sub-phase, referred to as $S_2$, is used to obtain the side information required for efficient learning. Specifically, the opportunistic CSMA is executed again, but the assigned users of each channel do not transmit. All other users that attempted to transmit in $S_1$ transmit again on the same channel $k$. The sample mean of the best user in $S_2$ (i.e., the second best user in $\mathcal{T}_k$ for each channel $k$) is stored by the assigned user. This sub-phase continues until all $M$ users in $S_2$ were observed, and the phase ends. \\
An example for $M=K=3$ is given next. The expected rate matrix is shown in Table \ref{table:1}. Table \ref{table:3} shows the transmission attempts made by the users in the allocation phase before the stable matching was achieved (the assigned users are shown in bold). At time $t=1$, each user transmits on its best channel (sub-phase $S_1$). Users $1$ and $2$ aim to access the same channel (channel $2$), and the channel is assigned to user $2$ since it has a higher expected rate on this channel (i.e., smaller backoff time). At time $t=2$, sub-phase $S_2$ is performed, in which user $1$ transmits again on channel $2$. At time $t=3$, user $1$ (the only unassigned user) tries to access its second best channel; i.e., channel $1$. However, the channel is assigned to user $3$ since it has a higher expected rate. The algorithm continues until the three users are assigned to the three channels.
\ifOneCol
    \begin{table} [htbp!]
    \parbox{.45\linewidth}{
    \centering
    \caption{Expected rate matrix [Mbps]}
    \label{table:1}
    \begin{tabular}{ |p{1cm}||p{1.5cm}|p{1.5cm}|p{1.5cm}|}
    \hline
    & channel 1 & channel 2 & channel 3\\
    \hline \hline
    user 1 &  45  & 70 & 35\\
    \hline
    user 2 &  30  & 90 & 60\\
    \hline
    user 3 &  65  & 10 & 50\\
    \hline
    \end{tabular}
    }
    \hfill
    \parbox{.45\linewidth}{
    \centering
    \caption{Allocation phase}
    \label{table:3}
    \begin{tabular}{|p{1cm}||p{1cm}||p{1.2cm}|p{1.2cm}|p{1.2cm}|}
    \hline
    Sub-phase & Time & channel 1 & channel 2 & channel 3\\
    \hline \hline
    $S_1$ & t=1 &  $\boldsymbol{3}$  & 1\;,\;$\boldsymbol{2}$ & \\
    \hline
    $S_2$ & t=2 &     & 1 & \\
    \hline
    $S_1$ & t=3 & 1\;,\;$\boldsymbol{3}$  & $\boldsymbol{2}$ & \\
    \hline
     $S_2$ & t=4 &  1  &  & \\
    \hline
    $S_1$ & t=5 &  $\boldsymbol{3}$  & $\boldsymbol{2}$ &  $\boldsymbol{1}$\\
    \hline
    \end{tabular}
    }
    \end{table}
\else
    \begin{table} [htbp!]
    \centering
    \caption{expected rate matrix}
    \label{table:1}
    \begin{tabular}{ |p{1cm}||p{1.5cm}|p{1.5cm}|p{1.5cm}|}
    \hline
    U& channel 1 & channel 2 & channel 3\\
    \hline \hline
    user 1 &  45  & 70 & 35\\
    \hline
    user 2 &  30  & 90 & 60\\
    \hline
    user 3 &  65  & 10 & 50\\
    \hline
    \end{tabular}
    \end{table}

    \begin{table} [htbp!]
    \centering
     \caption{allocation phase}
     \label{table:3}
     \begin{tabular}{|p{1cm}||p{1cm}||p{1.2cm}|p{1.2cm}|p{1.2cm}|  }
      \hline
      Sub-phase & Time & channel 1 & channel 2 & channel 3\\
      \hline \hline
     $S_1$ & t=1 &  3  & 1,2 & \\
      \hline
     $S_2$ & t=2 &     & 1 & \\
      \hline
     $S_1$ & t=3 & 1,3  & 2 & \\
      \hline
     $S_2$ & t=4 &  1  &  & \\
      \hline
     $S_1$ & t=5 &  3  & 2 &  1\\
     \hline
     \end{tabular}
    \end{table}
\fi

\subsection{The structure of the 
exploitation phase:}
\label{ssec:exploitation}
Let $n_I(t)$ be the number of exploitation phases up to time $t$. In the exploitation phase, each user transmits on the channel it was assigned according to the last allocation phase (during $S_1$) for a deterministic period of time $2 \cdot 4^{n_I(t)-1}$ (for the $n_I^{th}$ exploitation phase). There are no channel switching and no sample mean updating during the exploitation phase.

\subsection{Parameter setting for efficient learning:}
\label{ssec:parameter}
As discussed earlier, exploring the channels increases the regret since the stable matching allocation is not used. On the other hand, it is essential to reduce the estimation error and hence reduce the regret scaling order with time. In this section, we establish the sufficient exploration rate of each channel for each user to achieve efficient learning of the stable matching allocation. We next establish two parameters used in the learning strategy.

\subsubsection{Identifying $M$ best channels}
\label{sssec:identifying}
We show in the analysis that a user (say user $i$) who is interested in distinguishing with a sufficiently high accuracy between two channels $k,l$ that yield expected rates $\mu_{i,k},\mu_{i,\ell}$, respectively, must explore them at least $\displaystyle \frac{4L}{(\mu_{i, k}-\mu_{i,\ell})^2} \cdot \log(t)$ times.   
Let $\mathcal{M}_i$ be the set of the $M$ best channels of user $i$. For each channel $k \in \mathcal{M}_i$ we define the deterministic row\footnote{This definition is consistent with the definition of the $M \times K $ expected rate matrix by $U = \{\mu_{ik}\}$, $i=1, ..., M, k=1, ..., K$.} exploration coefficient as
\beq
\label{eq:Det_best_row}
\bea{l}
\displaystyle D_{i,k}^{(R)} \triangleq \frac{4L}{ \displaystyle \min_{\ell \neq k} \{ (\mu_{i,k}-\mu_{i,\ell})^2 \}}, 
\ena
\eeq 
and for channel $k \nin \mathcal{M}_i$,
\beq
\label{eq:Det_worst_row}
\bea{l}
\displaystyle D_{i,k}^{(R)} \triangleq \frac{4L}{(\mu_{i,k}-\mu_{i,\sigma_i(M)})^2 }. 
\ena
\eeq 
Since the expected rates are unknown, the users need to estimate $D_{i,k}^{(R)}$ for each channel $k \in \mathcal{K}$. This estimator is denoted by $\widehat{D}_{i,k}^{(R)}(t)$. Let $\bar{s}_{i,k}(t)$ be the mean transmission rate of user $i$ on channel $k$. Thus, the adaptive row exploration coefficient for channels $k \in \mathcal{M}_i$ is defined by
\beq
\label{eq:Est_best_row}
\bea{l}
\displaystyle \widehat{D}_{i,k}^{(R)}(t) \triangleq \frac{4L}{ \max \big\{ \Delta_{\min}^2, \displaystyle \min_{\ell \neq k} \{ (\bar{s}_{i,k}(t)-\bar{s}_{i,\ell}(t))^2 \} -\epsilon \big\} }, 
\ena
\eeq  
and similarly for $k \nin \mathcal{M}_i$ we have:
\beq
\label{eq:Est_worst_row}
\bea{l}
\displaystyle \widehat{D}_{i,k}^{(R)}(t) \triangleq \frac{4L}{ \max \{ \Delta_{\min}^2, (\bar{s}_{i,k}(t)-\bar{s}_{i,\sigma_i(M)}(t))^2-\epsilon\} }, \ena
\eeq  
where $\Delta_{\min}$ is the smallest difference between two entries in the expected rate matrix $U$; i.e., 
\ifOneCol
\begin{center}
$\hspace{3cm} \Delta_{\min} \triangleq \displaystyle \min_{i \in \mathcal{M}} \Delta_i \hspace{0.5cm}, \hspace{0.5cm} \Delta_i \triangleq \displaystyle \min_{k, \ell \in \mathcal{K}, k \neq \ell} |\mu_{i,k}-\mu_{i,\ell}|. $
\end{center}
\else
\begin{center}
$\hspace{3cm} \Delta_{\min} \triangleq \displaystyle \min_{i \in \mathcal{M}} \Delta_i$, \\
$\vspace{0.2cm} \hspace{2.5cm} \Delta_i \triangleq \displaystyle \min_{k, \ell \in \mathcal{K}, k \neq \ell} |\mu_{i,k}-\mu_{i,\ell}|. $
\end{center}
\fi

\subsubsection{CSMA protocol identification}       
Consistent with the opportunistic CSMA protocol described above, each user $i$ needs to distinguish between a channel $k \in \mathcal{T}_k$ (this channel is in  $\mathcal{M}_i$ as well), and the best channel in $\mathcal{T}_k$ (and the second best channel in $\mathcal{T}_k$ if $k$ is the best channel in $\mathcal{T}_k$), for all $k$. Hence, we define the deterministic column exploration coefficient for user $i$ for channel $k \in\mathcal{T}_k $ by: 
\beq
\label{eq:Det_col}
\bea{l}
\displaystyle D_{i,k}^{(C)} \triangleq \frac{4L}{ ( \mu_{i,k} - \displaystyle \max_{j \neq i, j \in \mathcal{T}_k} \mu_{j,k})^2 } ,
\ena
\eeq  
and the adaptive column exploration coefficient by:
\beq
\label{eq:Est_col}
\bea{l}
\displaystyle \widehat{D}_{i,k}^{(C)}(t) \triangleq \frac{4L}{\displaystyle \max \{ \Delta_{\min}^2, (\bar{s}_{i,k}(t) - \max_{j \neq i} \bar{s}_{j,k}(t))^2 -\epsilon  \} }. 
\ena
\eeq  
Note that $\max_{j \neq i, j \in \mathcal{T}_k} \bar{s}_{j,k}(t)$ is known to user $i$ by the design of the opportunistic CSMA (by sub-phase $S_2$). By combining (\ref{eq:Det_best_row}) and (\ref{eq:Det_col}), the deterministic exploration-rate coefficient of user $i$ for channels $k \in \mathcal{M}_i \cap \mathcal{T}_k$ is given by:
\beq
\label{eq:Det_best}
\bea{l}
\displaystyle D_{i,k} \triangleq \max \{ D_{i,k}^{(R)},D_{i,k}^{(C)} \},
\ena
\eeq  
and by combining (\ref{eq:Est_best_row}) and (\ref{eq:Est_col}), the adaptive exploration-rate coefficient of user $i$ for channels $k \in \mathcal{M}_i \cap \mathcal{T}_k$ is given by:
\beq
\label{eq:Est}
\bea{l}
\displaystyle \widehat{D}_{i,k}(t) = \max \{ \widehat{D}_{i,k}^{(R)}(t),\widehat{D}_{i,k}^{(C)}(t) \}.
\ena
\eeq  
\begin{remark}
The design of the adaptive exploration-rate coefficients under DSSL significantly reduces the regret as compared to existing algorithms that use deterministic exploration-rate coefficients determined by the channel that requires the largest exploration time \cite{nayyar2016regret,avner2016multi,bistritz2018distributed, 2019arXiv190201239B}. For example, consider the expected rate matrix $U$ given in Table \ref{table:1}, where parameter $L$ in (\ref{eq:L}) equals $10^4$. In Table \ref{table:4}, we present the deterministic exploration-rate coefficients $D_{i,k}$ defined in (\ref{eq:Det_best}) for each channel-user pair under DSSL, where $D_{i,k} \cdot \log(t)$ is the number of samples required to achieve consistent estimates of the expected rates. By contrast, in other existing algorithms \cite{nayyar2016regret,avner2016multi,bistritz2018distributed, 2019arXiv190201239B}, all channels are explored with the same exploration-rate coefficient, which is inversely proportional to the squared difference between the mean rate of the optimal allocation and the second best one. When applying this to our example, each channel should be explored for $1600 \cdot \log(t)$ time steps (as seen in Table \ref{table:5}), which significantly increases the exploration times unnecessarily, and consequently increases the regret. 
\end{remark}

\ifOneCol
    \begin{table} [htbp!]
    \parbox{.45\linewidth}{
    \centering
    \caption{Exploration coefficients under the DSSL algorithm}
    \label{table:4}
    \begin{tabular}{|p{1cm}|| p{1.5cm}| p{1.5cm}| p{1.5cm}|}
    \hline
    $D_{i,k}$& channel 1 & channel 2 & channel 3\\
    \hline \hline
    user 1 &  400  & 100 & 400\\
    \hline
    user 2 &  45  & 100 & 45\\
    \hline
    user 3 &  178  & 25 & 178\\
    \hline
    \end{tabular}
    }
    \hfill
    \parbox{.45\linewidth}{
    \centering
    \caption{Exploration coefficients under other existing algorithms \cite{nayyar2016regret,avner2016multi,bistritz2018distributed, 2019arXiv190201239B}}
    \label{table:5}
    \begin{tabular}{ |p{1cm}||p{1.5cm}|p{1.5cm}|p{1.5cm}|  }
    \hline
    $D_{i,k}$& channel 1 & channel 2 & channel 3\\
    \hline \hline
    user 1 &  1600  & 1600 & 1600\\
    \hline
    user 2 &  1600  & 1600 & 1600\\
    \hline
    user 3 &  1600  & 1600 & 1600\\
    \hline
    \end{tabular}
    }
    \end{table}

\else
    \begin{table} [htbp!]
    \centering
  \caption{Exploration coefficients under the DSSL algorithm}
 \label{table:4}
 \begin{tabular}{|p{1cm}|| p{1.5cm}| p{1.5cm}| p{1.5cm}|}
  \hline
  $D_{i,k}$& channel 1 & channel 2 & channel 3\\
  \hline \hline
  user 1 &  400  & 100 & 400\\
  \hline
  user 2 &  45  & 100 & 45\\
  \hline
  user 3 &  178  & 25 & 178\\
  \hline
 \end{tabular}
    \end{table}

    \begin{table} [htbp!]
    \centering
    \caption{Exploration coefficients under other existing algorithms \cite{nayyar2016regret,avner2016multi,bistritz2018distributed, 2019arXiv190201239B}}
    \label{table:5}
    \begin{tabular}{ |p{1cm}||p{1.5cm}|p{1.5cm}|p{1.5cm}|  }
  \hline
  $D_{i,k}$& channel 1 & channel 2 & channel 3\\
  \hline \hline
  user 1 &  1600  & 1600 & 1600\\
  \hline
  user 2 &  1600  & 1600 & 1600\\
  \hline
  user 3 &  1600  & 1600 & 1600\\
  \hline
 \end{tabular}
    \end{table}
\fi

\subsection{Choosing between phases types:}
\label{ssec:selection}
Since $D_{i,k}$ is unknown, the algorithm replaces $D_{i,k}$ by its estimate $\widehat{D}_{i,k}(t)$. Furthermore, to ensure that $\widehat{D}_{i,k}(t)$ overestimates $D_{i,k}$, the users need to sense at least $I \cdot \log (t)$ times each of their channels in exploration phases, where 
\beq
\label{eq:rate_function}
\bea{l}
\displaystyle I \triangleq \frac{7\epsilon^2 }{48(r_{\max}+2)^2 \cdot L},
\ena
\eeq  
which can be viewed as the rate function of the estimators among all channels. At the end of the exploitation phases, the users check the condition:
\beq
\label{eq:Condition}
\bea{l}
\displaystyle T_{i,k}^{(O)}(t)> \max \left\{\widehat{D}_{i,k}(t),\frac{2}{I} \right\} \cdot \log (t),
\ena
\eeq 
where $T_{i,k}^{(O)}(t)$ is the number of samples in the exploration phases accessed in sub epochs DE for user $i$ on channel $k$ up to time $t$.

If the condition holds for user $i$, the user enters another exploitation phase by transmitting on the same channel in which it transmitted during the last exploitation phase. Otherwise, if the condition does not hold, the user enters an exploration phase by sensing channel $k$. At the end of the phase, the user signals the other users that it has finished the exploration phase. If such an interruption occurred, all the users again check condition (\ref{eq:Condition}). If it holds for all users, they start an allocation phase. At the end of the allocation phase, an exploitation phase starts. A pseudocode of the DSSL algorithm is provided in Algorithm \ref{Algorithm}.

\begin{algorithm}
\footnotesize
\caption{DSSL Algorithm for user $i$}\label{Algorithm}
\begin{algorithmic}
\STATE \textbf{Initialization:} For all $K$ channels, execute an exploration phase where a single observation is taken from each channel;
\WHILE {$t \leq T$}
\IF {Condition (\ref{eq:Condition}) does not hold for channel $k$}
\STATE Enter an exploration phase with length $4^{n_O^{i,k}(t)}$;
\STATE Update $\bar{s}_{i,k}(t)$ and increment $n_O^{i,k}(t)=n_O^{i,k}(t)+1$;
\STATE \textbf{goto} step 3
\ENDIF
\STATE Send an interrupt signal; 
\STATE Start an allocation phase;
\STATE Start an exploitation phase with length $2 \cdot 4^{n_I(t)}$. If an interruption occurs, go to step $3$; 
\STATE $n_I(t) = n_I(t)+1$;
\ENDWHILE
\end{algorithmic}
\end{algorithm}

\section{Regret Analysis}

Success in obtaining a logarithmic regret order depends on how fast $\widehat{D}_{i,k}(t)$ converges to a value which is no smaller than $D_{i,k}$ (so that user $i$ senses channel $k$ at least $D_{i,k} \cdot \log t$ time slots in most of the times). The analysis in the Appendix shows that exploring channels as in (\ref{eq:Condition}) guarantees the desired convergence speed. Specifically, in the following theorem we establish a finite-sample bound on the regret with time, which results in a logarithmic scaling of the regret. 

\begin{theorem}
\label{th:regret}
Assume that the proposed DSSL algorithm is implemented and that the assumptions on the system model described in Section \ref{sec:problem} hold. Then, the regret at time $t$ is upper bounded by:
\ifOneCol
\beq
\bea{l}
\label{eq:total_regret}
\vspace{0.0cm} \hspace{0.3cm} \displaystyle  r(t) \leq A_{\max} \cdot \bigg(\sum\limits_{i=1}^{M} \sum\limits_{k=1}^{K} (\lfloor \log_4(3A_{i,k}\log(t)+1) \rfloor +1) \bigg) \displaystyle + \sum\limits_{i=1}^{M} \sum\limits_{k=1}^{K} \bigg[\bigg( 4A_{i,k} \cdot \log(t) + 1 \\ 
\vspace{0.0cm} \hspace{1.5cm} \displaystyle + M_{\max}^{i,k}\big(\lfloor \log_4(3A_{i,k}\log(t)+1) \rfloor +1 \big) \bigg) \displaystyle \cdot \bigg( \mu_{i,S(i)} + \mu_{S^{-1}(k),k} - \mu_{i,k} \bigg) \bigg] \\
\vspace{0.0cm} \hspace{0.7cm} \displaystyle + M^2 \cdot A_{\max} \cdot \bigg(\sum\limits_{i=1}^{M} \sum\limits_{k=1}^{K} (\lfloor \log_4(3A_{i,k}\log(t)+1) \rfloor +1) \bigg) \\
\vspace{0.0cm} \hspace{0.7cm} \displaystyle + \bigg[\bigg(2e \cdot \log(M+1) \bigg) \cdot \bigg(\sum\limits_{i=1}^{M} \sum\limits_{k=1}^{K} (\lfloor \log_4(3A_{i,k}\log(t)+1) \rfloor +1) \bigg)  \bigg] \displaystyle \cdot \bigg[\sum\limits_{j=1}^{M} \mu_{j,S(j)} \bigg] \\
\vspace{0.0cm} \hspace{0.7cm} \displaystyle + \bigg(A_{\max} + (M^2K+MK) \frac{6X_{\max}}{\pi_{\min}}\big(\sum \limits_{j=1}^M \mu_{j,S(j)}\big) \bigg)  \displaystyle \cdot \bigg(\lceil \log_4( \frac{3}{2}t+1) \rceil \bigg) +O(1),
\ena
\eeq
\else
\begin{center}
$\bea{l}
\label{eq:total_regret}
\vspace{0.3cm} \hspace{0.3cm} \displaystyle  r(t) \leq A_{\max} \cdot \bigg(\sum\limits_{i=1}^{M} \sum\limits_{k=1}^{K} (\lfloor \log_4(3A_{i,k}\log(t)+1) \rfloor +1) \bigg) \\
\vspace{0.3cm} \hspace{0.7cm} \displaystyle + \sum\limits_{i=1}^{M} \sum\limits_{k=1}^{K} \bigg[\bigg( 4A_{i,k} \cdot \log(t) + 1 \\ 
\vspace{0.3cm} \hspace{1.5cm} \displaystyle + M_{\max}^{i,k}\big(\lfloor \log_4(3A_{i,k}\log(t)+1) \rfloor +1 \big) \bigg) \\
\vspace{0.3cm} \hspace{3.5cm} \displaystyle \cdot \bigg( \mu_{i,S(i)} + \mu_{S^{-1}(k),k} - \mu_{i,k} \bigg) \bigg] \ena$    
\end{center}
\beq
\bea{l}
\label{eq:total_regret}
\vspace{0.3cm} \hspace{0.7cm} \displaystyle + M^2 \cdot A_{\max} \cdot \bigg(\sum\limits_{i=1}^{M} \sum\limits_{k=1}^{K} (\lfloor \log_4(3A_{i,k}\log(t)+1) \rfloor +1) \bigg) \\
\vspace{0.3cm} \hspace{0.7cm} \displaystyle + \bigg[\bigg(2e \log(M+1) \bigg) \\
\vspace{0.3cm} \hspace{1.2cm} \cdot \bigg(\sum\limits_{i=1}^{M} \sum\limits_{k=1}^{K} (\lfloor \log_4(3A_{i,k}\log(t)+1) \rfloor +1) \bigg)  \bigg] \\
\vspace{0.3cm} \hspace{0.8cm} \displaystyle \cdot \bigg[\sum\limits_{j=1}^{M} \mu_{j,S(j)} \bigg] \\
\vspace{0.3cm} \hspace{0.7cm} \displaystyle + \bigg(A_{\max} + (M^2K+MK) \frac{6X_{\max}}{\pi_{\min}}\big(\sum \limits_{j=1}^M \mu_{j,S(j)}\big) \bigg) \\
\vspace{0.3cm} \hspace{0.8cm} \displaystyle \cdot \bigg(\lceil \log_4( \frac{3}{2}t+1) \rceil \bigg) +O(1),
\ena
\eeq
\fi

where $A_{i,k}$ is given by: 
\begin{align}
\vspace{0.0cm} A_{i,k}\triangleq
\left\{ \begin{matrix}
\max\{2/I\;,\;D_{i,k}^{(\max)} \} \;, & \mbox{if $k\in\mathcal{G}_i$}    \vspace{0.0cm} \\
\max \{2/I\;,\;4L/\Delta_{\min}^2\} \;, & \mbox{if $k\nin\mathcal{G}_i$}
\end{matrix} \right. \;, \label{eq:20}
\end{align}

$\mathcal{G}_i$ is defined as the set of all indices $k\in \mathcal{K}$ of user $i$ that satisfy:\vspace{0.0cm}
\begin{center}
$\displaystyle \min \{(\displaystyle \min_{\ell \neq k} \{ \mu_{i,k}-\mu_{i,\ell} \})^2, (\mu_{i,k}-\max_{j \neq i} \mu_{j,k})^2 \} -2\epsilon > \Delta_{\min}^2, $ \vspace{0.0cm}
\end{center}
for $k \in \mathcal{T}_k$,
and \vspace{0.0cm}
\begin{center}
$\displaystyle (\displaystyle \min_{\ell \neq k} \{ \mu_{i,k}-\mu_{i,\ell} \})^2 -2\epsilon > \Delta_{\min}^2, $ \vspace{0.0cm}
\end{center}
for $k \nin \mathcal{T}_k$,
where $D_{i,k}^{(\max)}$ is defined as: 
\beq
\label{eq:D_max}
\bea{l}
\hspace{-0.2cm} \displaystyle D_{i,k}^{(\max)}\triangleq{\frac{4L}{\min \big\{(\displaystyle \min_{\ell \neq k} \{ \mu_{i,k}-\mu_{i,\ell} \})^2, (\mu_{i,k}-\max_{j \neq i} \mu_{j,k})^2 \big\} -2\epsilon}}. \vspace{0.0cm}
\ena
\eeq 
\noindent
\end{theorem}

The proof is given in the Appendix.

Note that Theorem \ref{th:regret} shows that similar to \cite{Liu_2013_Learning, nayyar2016regret,bistritz2018distributed,2019arXiv190201239B}, the regret under DSSL has a logarithmic order with time. DSSL, however, achieves this under the more general restless Markovian model, and also has significantly better scaling with $M,K$ and $\Delta_{\min}$. Specifically, under a common benchmark setting of equal rates among users (but still vary among channels), and $K>M$, which allows a theoretical comparison of learning efficiency between different algorithms, in \cite{nayyar2016regret} and \cite{Liu_2013_Learning} the regret scales as $O(\frac{MK}{(\Delta_{\min})^2}\log(t))$ ,in \cite{2019arXiv190201239B} as $O(\frac{M^3K}{(\Delta_{\min})^2}\log(t))$ and in \cite{bistritz2018distributed} the regret scales as $O(\frac{MK^2}{(\Delta_{\min})^2}\log(t))$. In contrast, under DSSL, the regret scales as $O((\frac{1}{(\Delta_{\min})^2}+MK)\log(t))$ due to the novel algorithm design that explores every channel according to its unique adaptive exploration rate, while guaranteeing efficient learning.

\section{Simulation Results}
\label{sec:Simulations}

In this section we present simulation results to evaluate the performance of DSSL numerically. In Subsection \ref{ssec:unknown_vs_known} we start by evaluating the convergence of DSSL under unknown restless fading FSMCs with respect to the stable matching solution solved under known restless fading FSMCs. We also evaluate the performance as compared to random allocation and the optimal centralized allocation schemes. Then, in Section \ref{ssec:unknown_vs_unknown} we examine the learning efficiency of DSSL as compared to other online learning algorithms under unknown restless FSMC, and verify our theoretical logarithmic regret. We performed $1,000$ Monte-Carlo experiments and averaged the performance over the experiments.

\subsection{Convergence of DSSL to stable matching}
\label{ssec:unknown_vs_known}

We start by describing the wireless channel model used in the simulations. Each user experiences a block fading channel which remains constant during each time slot, and varies between time slots. The channel response experienced by user $i$ at time slot $t$ is given by $h(i,t) = r(i,t) e^{j \rho (i,t)}$, where $r(i,t) = |h(i,t)|$ denotes the channel rate, and $\rho (i,t)$ denotes the channel phase experienced by user $i$ at time $t$.
Let $f(i,r)$ denote the Probability Density Function (PDF) of the fading channel rate $r(i)$ experienced by user $i$ (e.g., Rayleigh fading distribution in the simulations). We consider independent but non-identically distributed channels across users, and Markovian correlated channels across time slots. The FSMC model \cite{wang1995finite,sadeghi2008finite} partitions the range of the channel gain values into a finite number of intervals and represents each interval as a state of a Markov chain. The thresholds of the intervals at user $i$ are denoted by $\tau_n(i), n = 0, \ldots N$, where $0 = \tau_0(i) < \tau_1(i) < \ldots < \tau_{N-1}(i) < \tau_N(i) = \infty$.
The channel rate $r(i,t)$ experienced by user $i$ is said to be in state $g_n(i), 1<n<N$, if it lies in the interval: $t_{n-1}(i) \leq r(i,t) < \tau_n(i)$. The states are partitioned to yield an equal initial state probability for all states: 
\ifOneCol
   $\displaystyle \int_{\tau_{n-1}(i)}^{\tau_{n}(i)} f(i,r) dr = \frac{1}{N}, n=1, \ldots, N$. \\
\else
\begin{center}
   $\displaystyle \int_{\tau_{n-1}(i)}^{\tau_{n}(i)} f(i,r) dr = \frac{1}{N}, n=1, \ldots, N$. \\
\end{center}
\fi
The transition probability to transition from state $g_n(i)$ to state $g_\ell(i)$ is defined by:
\ifOneCol
\begin{center}
    $\displaystyle p_{n,\ell}(i) \triangleq Pr(\tau_{\ell-1}(i) \leq r(i,t+1) < \tau_\ell (i) | \tau_{n-1}(i) \leq r(i,t) < \tau_n (i) )$,
\end{center}
\else
\begin{center}
    $\displaystyle p_{n,\ell}(i) \triangleq Pr(\tau_{\ell-1}(i) \leq r(i,t+1) < \tau_\ell (i)$ \\ $| \tau_{n-1}(i) \leq r(i,t) < \tau_n (i) ) $
\end{center}
\fi
where $r(i,t)$ and $r(i,t+1)$ are the current channel gain and the channel gain in the next time slot experienced by user $i$, respectively. In the simulations, we quantized the channel gain to $6$ states; i.e., $N=6$, and we simulated a case of $3$ users and $5$ channels. The transition probability matrix $P$ and the expected rate matrix $U$ are given by: \\
\ifOneCol
\[  \vspace{0.0cm}  P = \left( \begin{array}{cccccc} 3/6 & 2/6 & 1/6 & 0 & 0 & 0 \\
                                  2/8 & 3/8 & 2/8 & 1/8 & 0 & 0 \\
                                  1/9 & 2/9 & 3/9 & 2/9 & 1/9 & 0 \\
                                  0 & 1/9 & 2/9 & 3/9 & 2/9 & 1/9 \\
                                  0 & 0 & 1/8 & 2/8 & 3/8 & 2/8 \\
                                  0 & 0 & 0 & 1/6 & 2/6 & 3/6
    \end{array} \right), \qquad
    U = \left( \begin{array}{ccccc}  45 & 70 & 35 & 17.5 & 12.5 \\
                                  27.5 & 90 & 60 & 15 & 20 \\
                                  65 & 10 & 50 & 16.5 & 30 
    \end{array} \right).  \]
\else
\begin{center}
   $ P = \left( \begin{array}{cccccc} 3/6 & 2/6 & 1/6 & 0 & 0 & 0 \\
                                  2/8 & 3/8 & 2/8 & 1/8 & 0 & 0 \\
                                  1/9 & 2/9 & 3/9 & 2/9 & 1/9 & 0 \\
                                  0 & 1/9 & 2/9 & 3/9 & 2/9 & 1/9 \\
                                  0 & 0 & 1/8 & 2/8 & 3/8 & 2/8 \\
                                  0 & 0 & 0 & 1/6 & 2/6 & 3/6
    \end{array} \right) $,
\end{center}

\begin{center}
   $ U = \left( \begin{array}{ccccc}  45 & 70 & 35 & 17.5 & 12.5 \\
                                  27.5 & 90 & 60 & 15 & 20 \\
                                  65 & 10 & 50 & 16.5 & 30 
    \end{array} \right) $.
\end{center}
\fi
 We compared the expected rate evolution of DSSL under unknown FSMCs against stable matching, random allocation and the optimal centralized allocation solved under known FSMCs. The optimal centralized algorithm served as an upper bound benchmark for all algorithms, and the stable matching served as an upper bound for DSSL. In the random allocation scheme users access an arbitrary channel with equal probability. As shown in Fig. \ref{fig:rates} the average rate under DSSL converged to that of the stable matching, as desired. The stable matching allocation allocates user 1 to channel 3, user 2 to channel 2, and user 3 to channel 1. Fig. \ref{fig:users} shows that the average achievable rate of each user in the DSSL algorithm converged to the stable allocation.
 
 \ifOneCol
  \begin{figure}[htbp]
  \centering
  \begin{minipage}{.5\textwidth}
 \includegraphics[width=\linewidth]{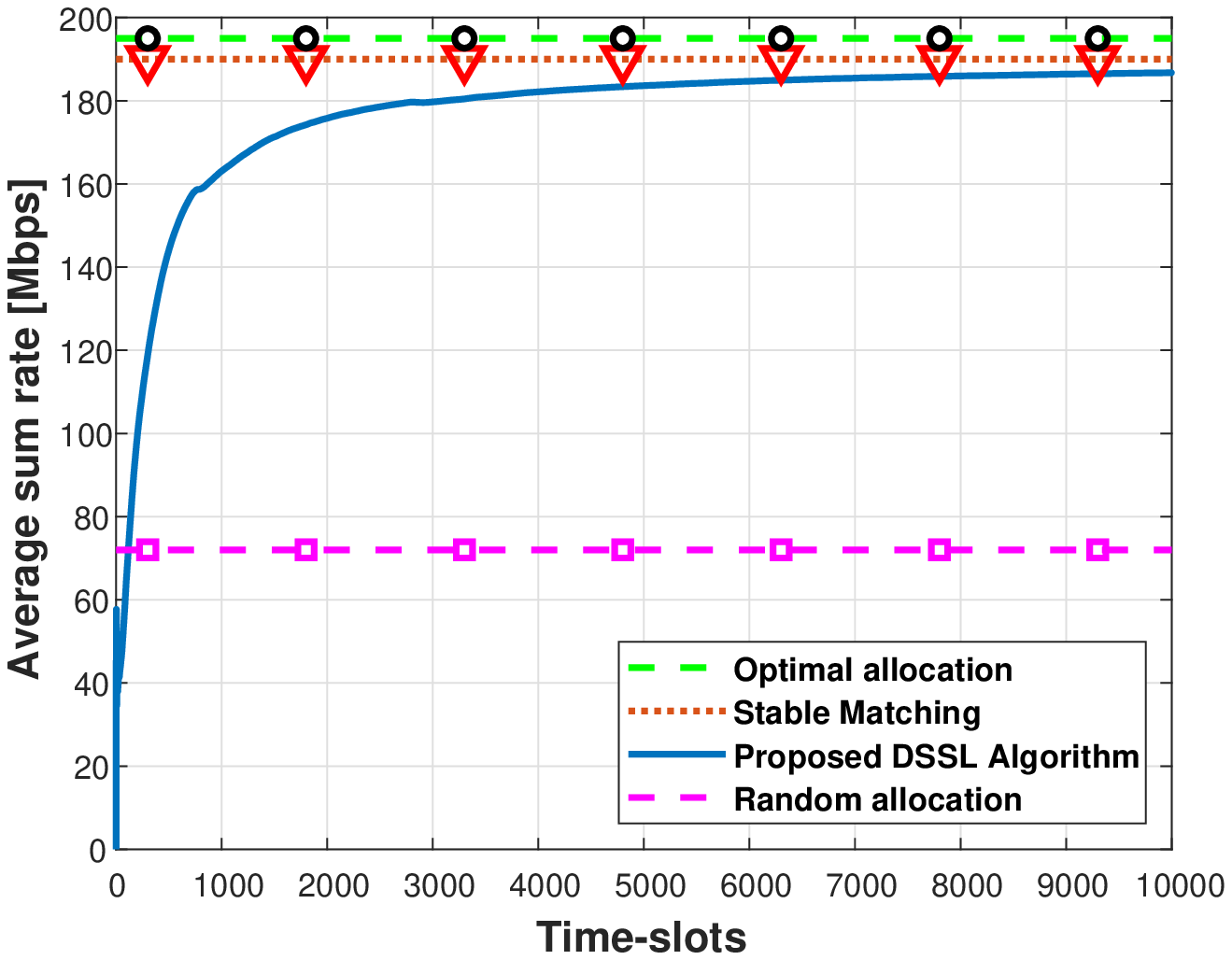}
 \captionof{figure}{Comparison of the system average rate of various schemes. } 
\label{fig:rates}
\end{minipage}%
\begin{minipage}{.5\textwidth}
\centering 
\includegraphics[width=\linewidth]{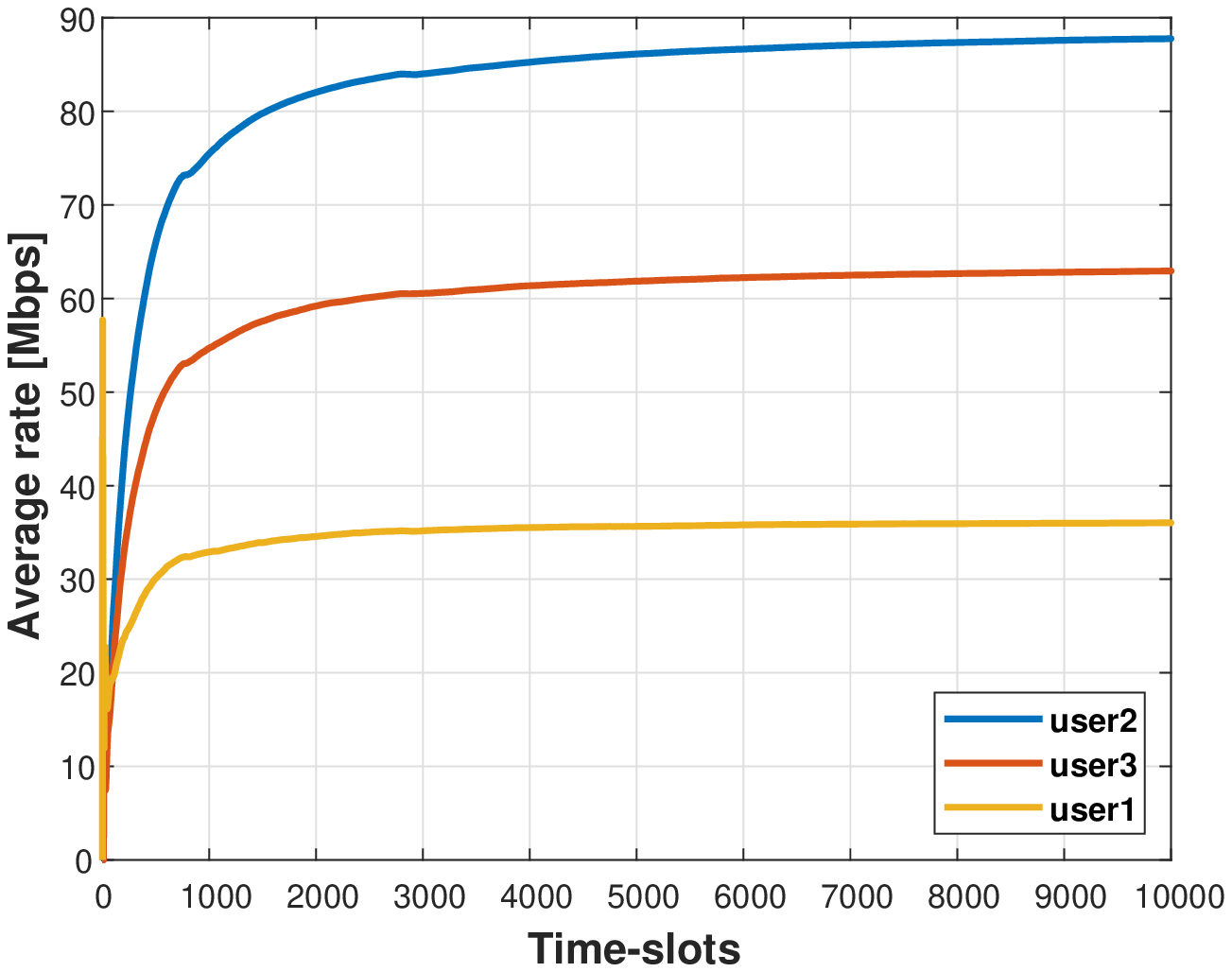}
\captionof{figure}{Comparison of users' average rate for the proposed DSSL algorithm.}
\label{fig:users}
\end{minipage}
\end{figure}
 \else
 \begin{figure}[htbp]
 \centering \epsfig{file=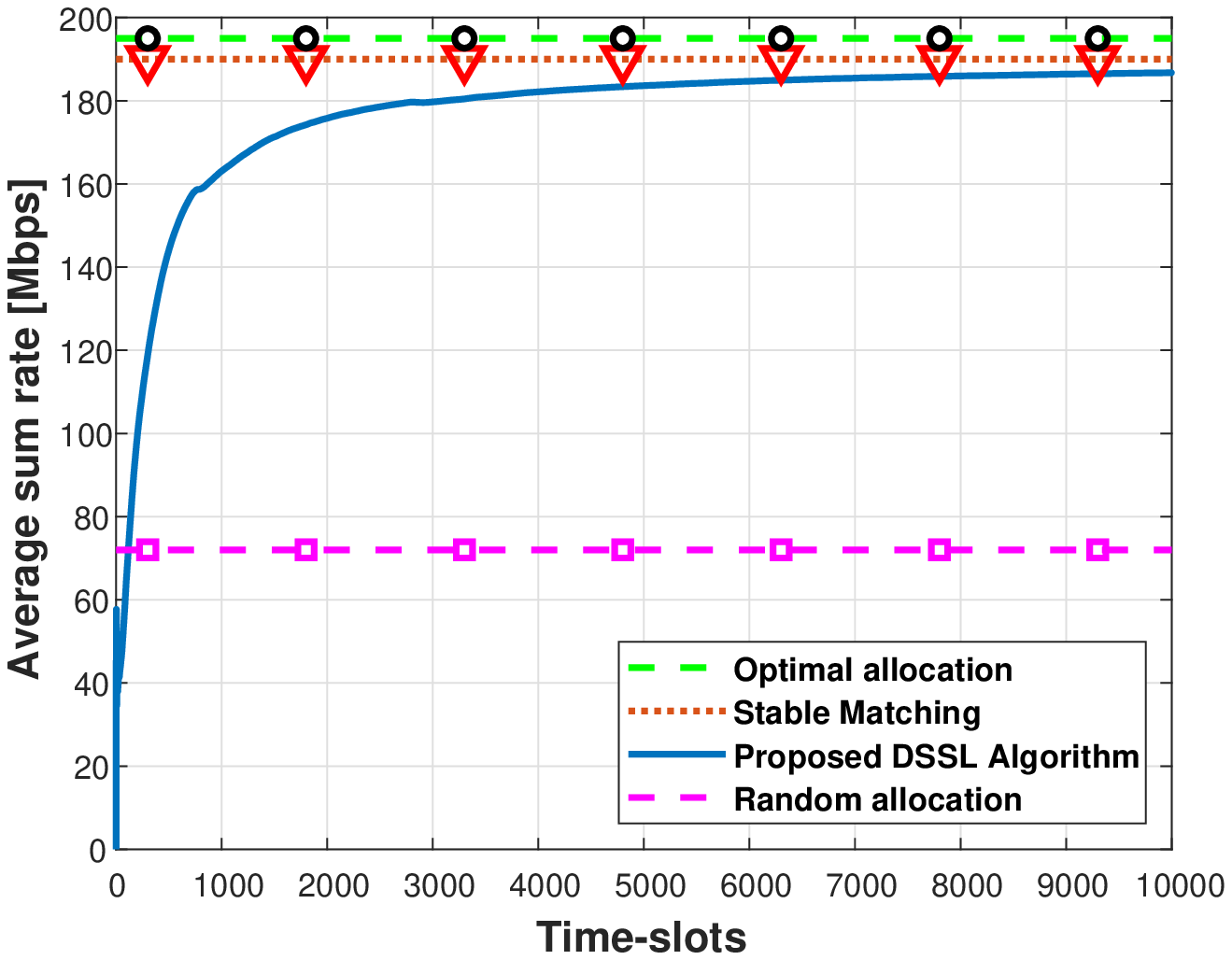,
width=0.45\textwidth}
\caption{Comparison of the system average rate of various schemes}
\label{fig:rates}
\end{figure}

\begin{figure}[htbp]
\centering \epsfig{file=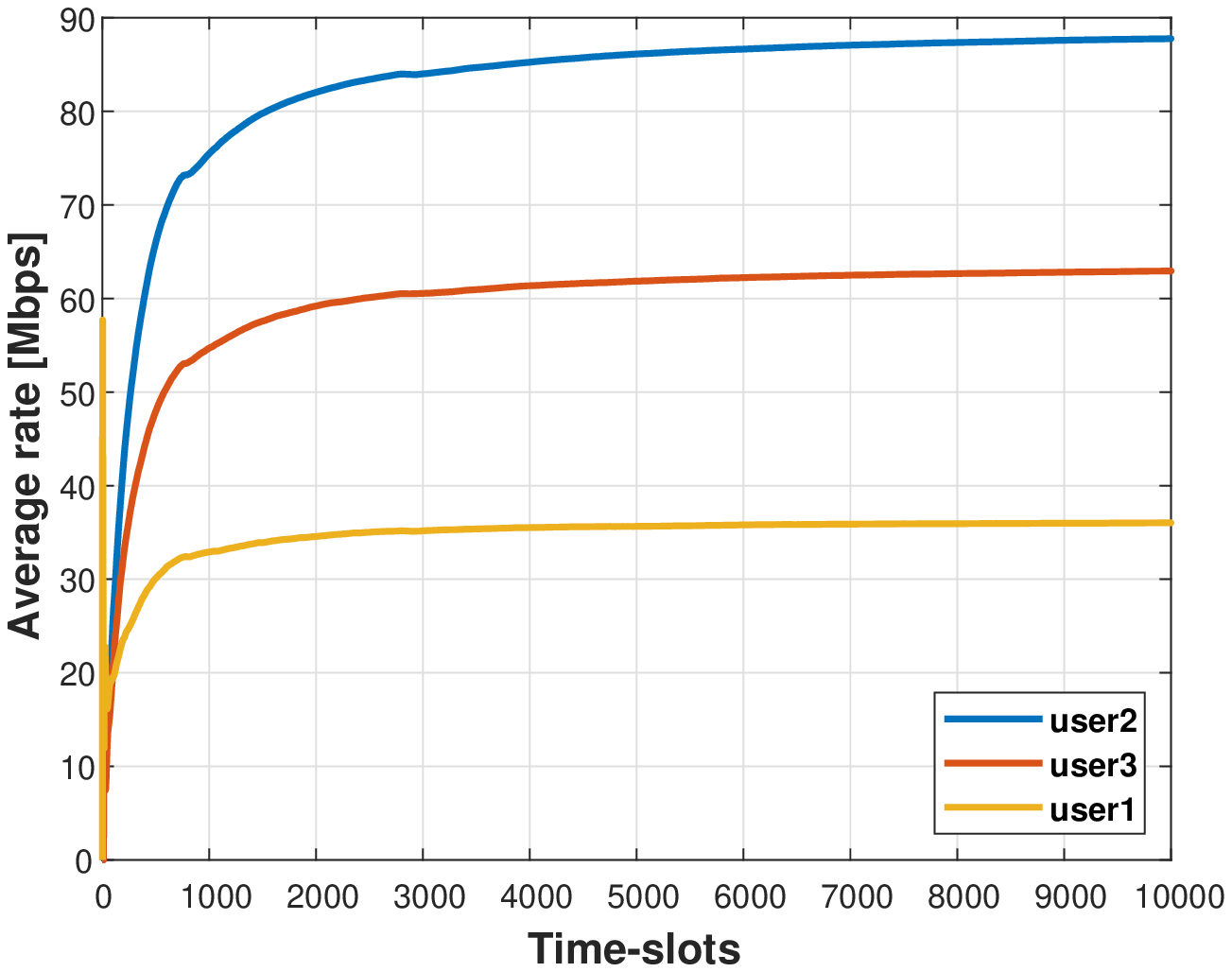,
width=0.45\textwidth}
\caption{Comparison of users' average rate for the proposed DSSL algorithm}
\label{fig:users}
\end{figure}
\fi

\subsection{Learning efficiency of DSSL}
\label{ssec:unknown_vs_unknown}

We next evaluated the learning efficiency of DSSL as compared to other online learning algorithms under unknown restless FSMCs. We considered the hierarchical access channel model in spectrum access networks. This models the situation of primary and secondary users that share the spectrum. Primary users (licensed) occupy the spectrum occasionally, and a secondary user is allowed to transmit over a single channel when the channel is free. Thus, each channel has two states, \textit{good} (free) and \textit{bad} (occupied). The good state results in a positive expected rate, whereas bad state result in a zero rate. The occupancies of the channels by the primary users are modeled as Markov processes (i.e., Gilbert-Elliott channel). 

First, we simulated a special case of our model where each channel yielded the same expected rate for all users. In \cite{Tekin_2012_Online,liu2012learning}, the RCA and DSEE algorithms were proposed to solve this special case. The RCA algorithm performs random regenerative cycles until catching predefined states in each phase, which results in oversampling the channels, and therefore is expected to increase the regret as compared to DSSL. The DSEE algorithm overcomes this issue by performing deterministic sequencing for both the exploration and exploitation phases. However, the deterministic sequencing requires the algorithm to explore all channels using the maximal exploration rate among all channels, which is expected to increase the regret as compared to DSSL (that learns the desired exploration rate for each channel) as well. We simulated the case of $2$ users, $6$ channels, each with two states: 0, 1. The transition probabilities for all channels to transition from 0 to 1 and from 1 to 0, respectively, were $p_{01} = [0.1, 0.1, 0.5, 0.1, 0.1,0.7]$, $p_{10} = [0.2, 0.3, 0.1, 0.4, 0.5,0.08]$, the expected rates for all channels at states 1, 0, respectively, are $r_1 = [1, 1, 1, 1, 1, 1]$, $r_0 =[0.1, 0.1, 0.1, 0.1, 0.1, 0.1]$. As can be seen in Fig. \ref{fig:special}, the DSSL algorithm outperformed both RCA and DSEE and achieved the logarithmic regret order with time.

Finally, we simulated the scenario where the stable matching allocation was also the optimal centralized allocation, and the channels were i.i.d. across time slots (and not Markovian). We compared DSSL to the $dE^3$ algorithm which was designed for this setting. However, $dE^3$ requires communication between users since it implements a distributed auction that requires users to observe the bids of other users \cite{nayyar2016regret}. We used the same parameters as selected and tuned  by the authors in \cite{nayyar2016regret}. Similar to the DSEE algorithm, in $dE^3$ the exploration-rate coefficient was determined by the channel with the largest exploration time. Thus, we expected that DSSL would yield a faster convergence rate due to the adaptive design of the exploration epochs. As shown in Fig. \ref{fig:fig3}, DSSL indeed outperformed the $dE^3$ algorithm.

\ifOneCol
  \begin{figure}[htbp]
  \centering
  \begin{minipage}{.5\textwidth}
 \includegraphics[width=\linewidth]{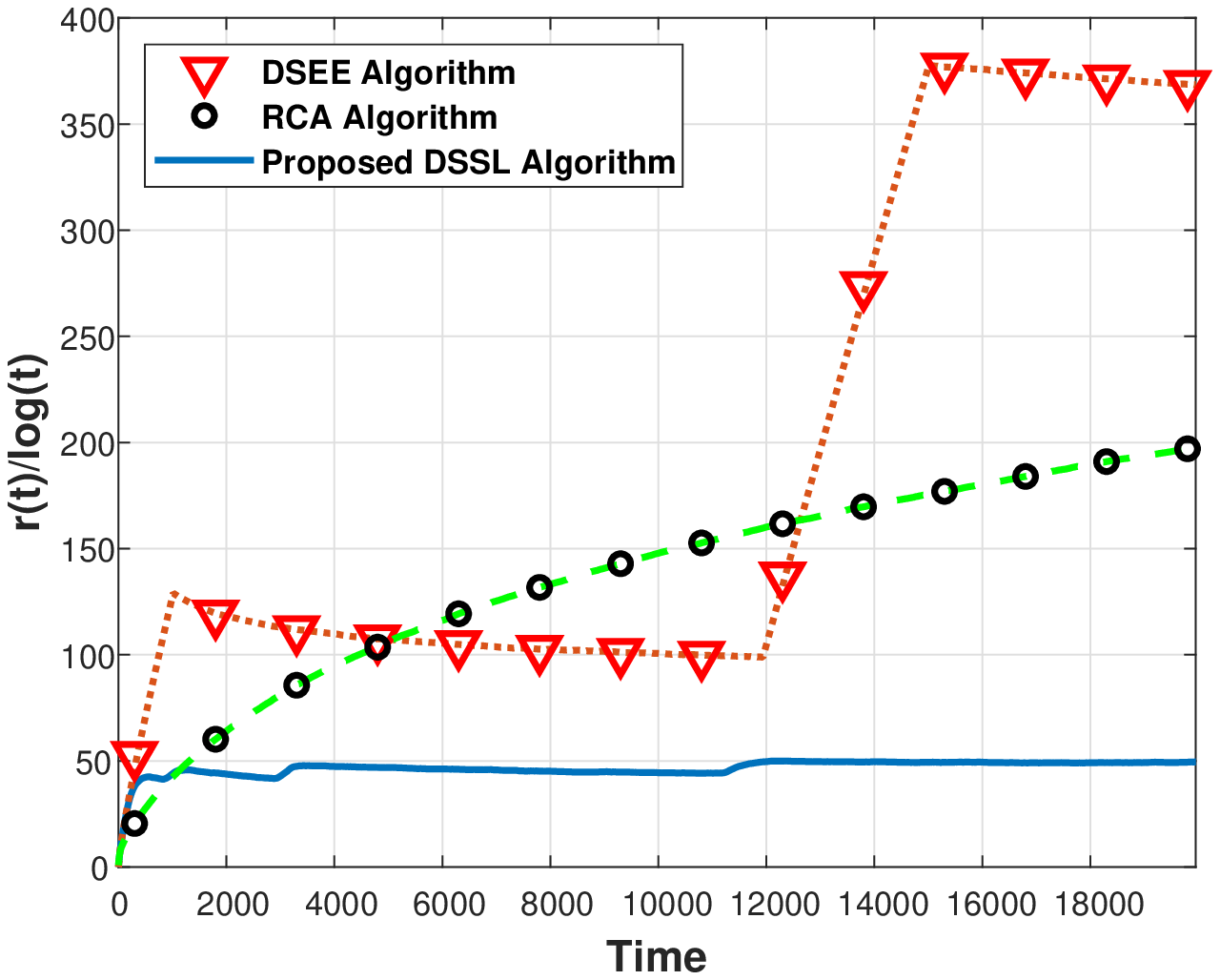}
\captionof{figure}{The regret under DSSL, DSEE, and RCA as a function of time.}
\label{fig:special}
\end{minipage}%
\begin{minipage}{.5\textwidth}
\centering 
\includegraphics[width=\linewidth]{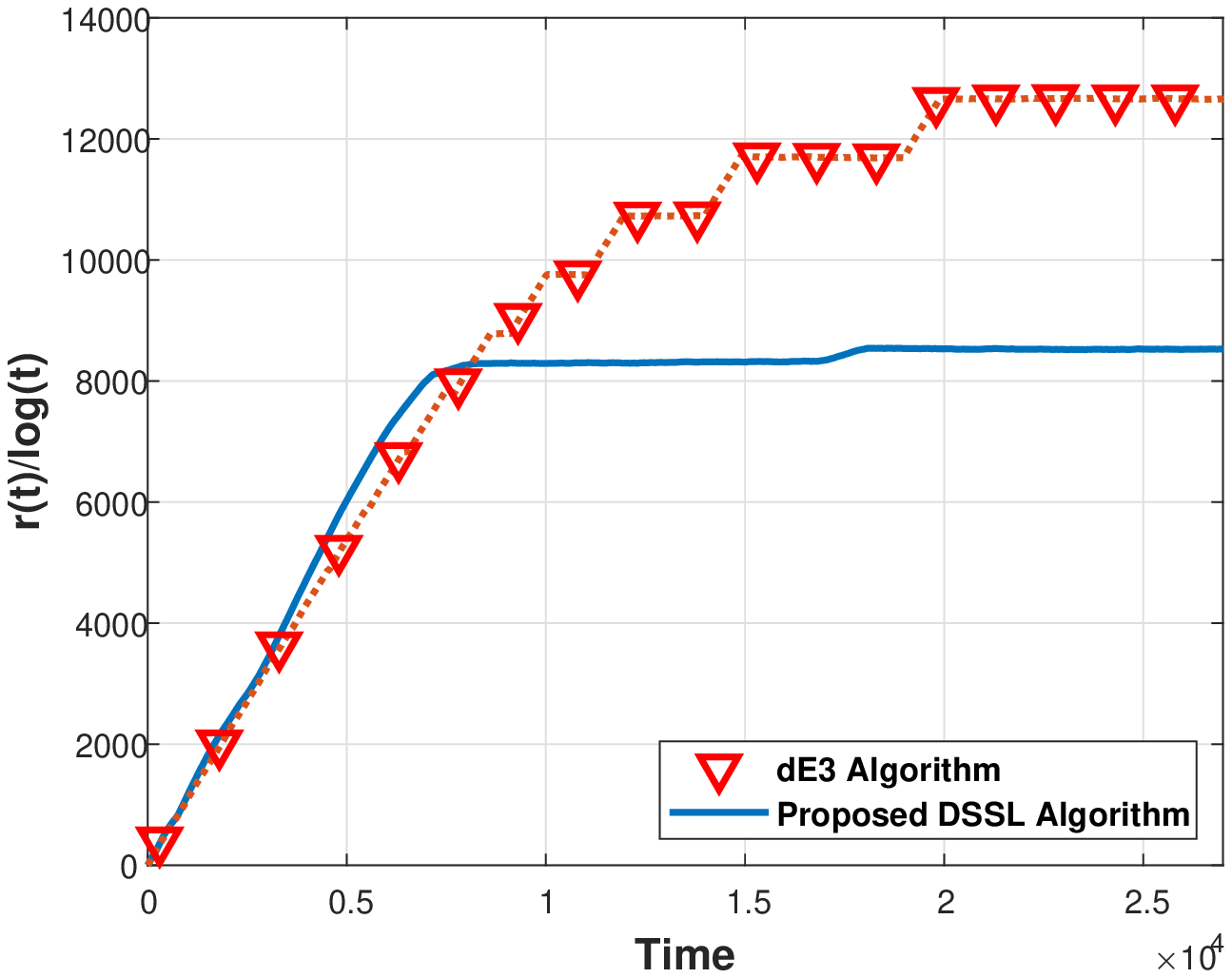}
\captionof{figure}{The regret under DSSL and $dE^3$ as a function of time.} 
\label{fig:fig3}
\end{minipage}
\end{figure}
\else
\begin{figure}[htbp]
\centering \epsfig{file=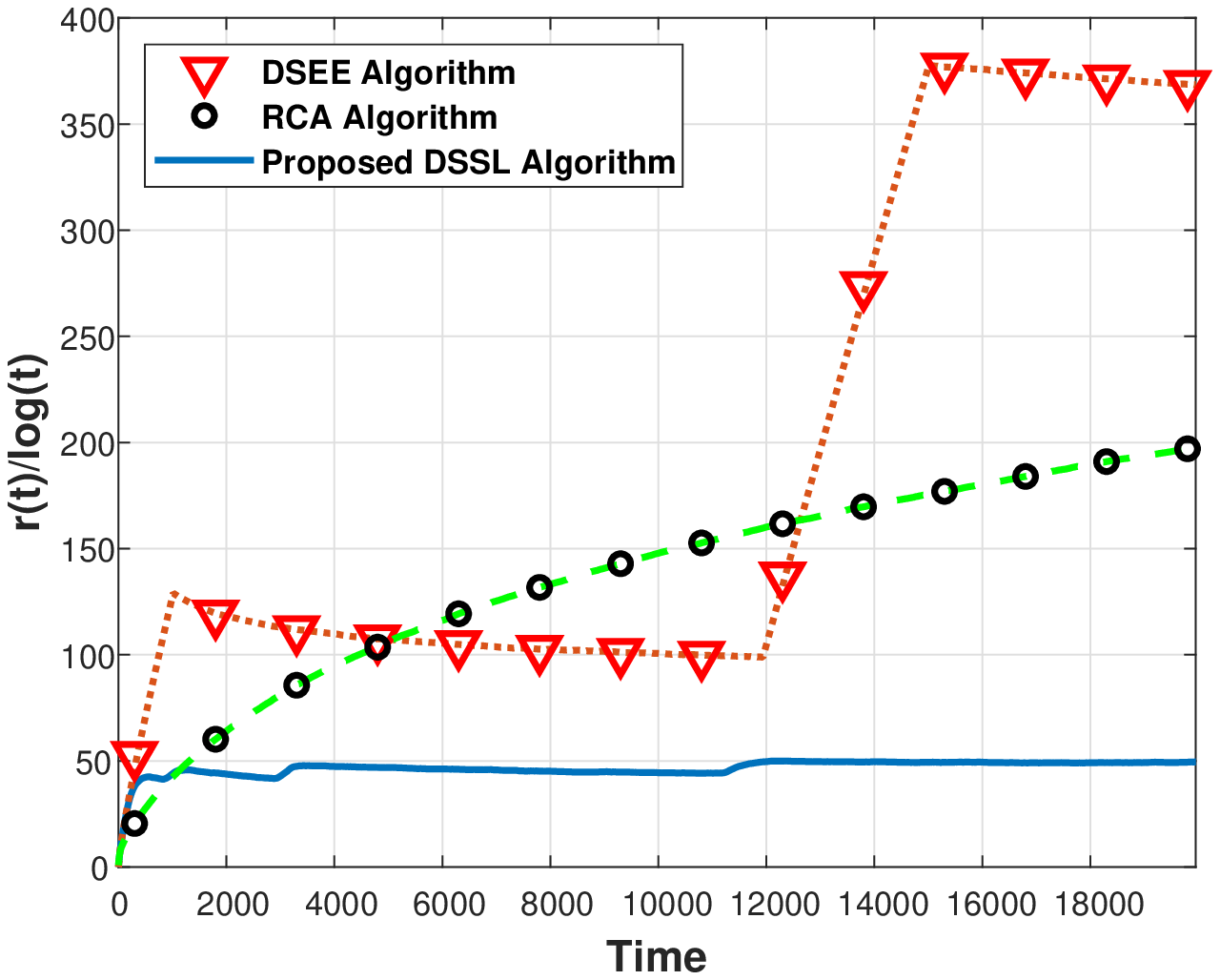,
width=0.45\textwidth}
\caption{The regret (normalized by $\log t$) under DSSL, DSEE, and RCA as a function of time. Parameter setting: 2 users, 6 channels, each with two states: 0, 1. Transition probabilities for all channels to transition from 0 to 1 and from 1 to 0, respectively: $p_{01} = [0.1, 0.1, 0.5, 0.1, 0.1,0.7]$, $p_{10} = [0.2, 0.3, 0.1, 0.4, 0.5,0.08]$, expected rates for all channels at states 1, 0, respectively: $r_1 = [1, 1, 1, 1, 1, 1]$, $r_0 =[0.1, 0.1, 0.1, 0.1, 0.1, 0.1]$.}
\label{fig:special}
\end{figure}

\begin{figure}[htbp]
\centering \epsfig{file=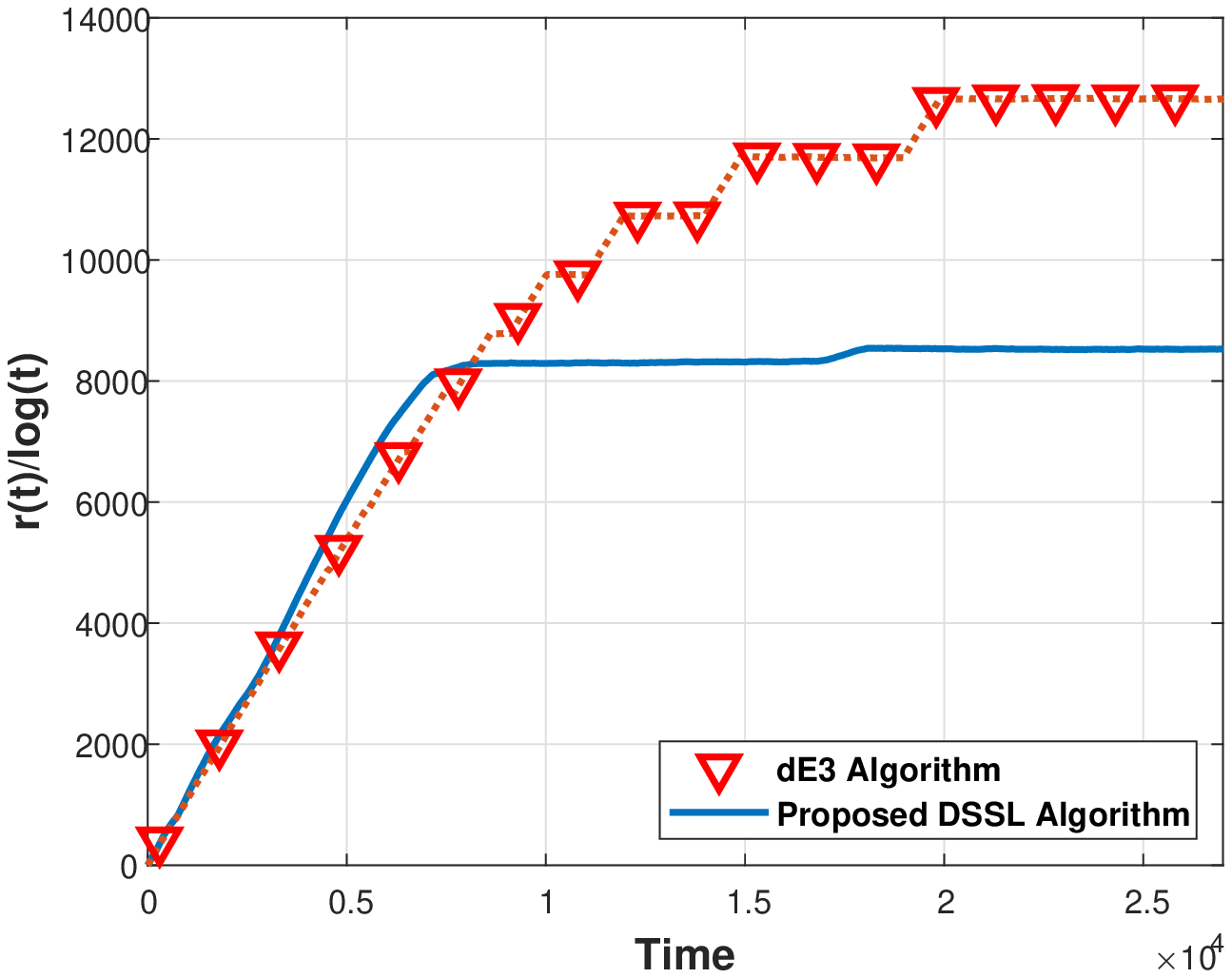,
width=0.45\textwidth}
\caption{The regret under DSSL and $dE^3$ as a function of time. Parameter setting: 3 users, 3 channels, with mean transmission rates: $[0.2,0.25,0.3; 0.4,0.6,0.5; 0.7,0.9,0.8]$.
}
\label{fig:fig3}
\end{figure}
\fi

\section{Conclusion}\label{sec:conclusion}
We developed a novel algorithm for the multi-user spectrum access problem in wireless networks, dubbed the Distributed Stable Strategy Learning (DSSL) algorithm. In contrast to existing models, for the first time we considered the case of restless Markov channels, which requires a different algorithm structure to accurately learn the channel statistics. Moreover, the channels selection rules are adaptive in order to reduce the exploration time required for efficient learning. We showed theoretically that DSSL achieves a logarithmic regret with time, and better regret scaling with the system parameters as compared to existing approaches that have studied special cases of the model. Extensive simulation results supported the theoretical study and demonstrated the strong performance of DSSL.     

\section{Appendix}\label{sec:appendix}

In this appendix we prove Theorem \ref{th:regret}. 

\begin{definition}
Let $T_1$ be the smallest integer, such that for all $t \geq T_1$ the following holds: $D_{i,k} \leq \widehat{D}_{i,k}(t)$ for all $i \in \mathcal{M}, k \in \mathcal{K}$, and also $\widehat{D}_{i,k}(t) \leq D_{i,k}^{(\max)}$ for all $i\in\mathcal{M}, k \in \mathcal{G}_i$.
\end{definition}

\begin{lemma}
\label{lemma:T1}
Assume that the DSSL algorithm is implemented as described in Section \ref{sec:DMASR}. Then, $E(T_1)<\infty$ is bounded independent of $t$. \vspace{0.0cm}
\end{lemma}

\textit{Proof}: $E(T_1)$ can be written as follows: \\

\ifOneCol
    $ E[T_1]=\sum\limits_{n=1}^{\infty} n \cdot Pr\left(T_1= n \right)=
    \sum\limits_{n=1}^{\infty}\Pr\left(T_1\geq n \right)= \\
    \vspace{0.0cm}
    =\sum\limits_{n=1}^{\infty} \Pr\big(\bigcup\limits_{i\in\mathcal{M}} \bigcup\limits_{k\in\mathcal{G}_i} \bigcup\limits_{l=n}^{\infty}(\widehat{D}_{i,k}(l)<D_{i,k} \mbox{\;or\;} \widehat{D}_{i,k}(l)> D_{i,k}^{(\max)}) \mbox{\;or\;} 
    \bigcup\limits_{i\in\mathcal{M}} \bigcup\limits_{k\nin\mathcal{G}_i}\bigcup\limits_{l=n}^{\infty}(\widehat{D}_{i,k}(l)<D_{i,k})
    \big)
    \\\leq
    \vspace{0.0cm} \hspace{0.0cm} \sum\limits_{i\in\mathcal{M}} \sum\limits_{k\in\mathcal{G}_i}\sum\limits_{n=1}^{\infty} \sum\limits_{l=n}^{\infty}
    \Pr\big(\widehat{D}_{i,k}(l)<D_{i,k} \mbox{\;or\;}  \widehat{D}_{i,k}(l)> D_{i,k}^{(\max)}\big)  +\sum\limits_{i\in\mathcal{M}}     \sum\limits_{k\nin\mathcal{G}_i}\sum\limits_{n=1}^{\infty}     \sum\limits_{l=n}^{\infty}
    \Pr\big(\widehat{D}_{i,k}(l)<D_{i,k}\big). $\\
\else
    $ E[T_1]=\sum\limits_{n=1}^{\infty} n \cdot Pr\left(T_1= n \right)=
    \sum\limits_{n=1}^{\infty}\Pr\left(T_1\geq n \right)\\=
    \vspace{0.0cm} \hspace{0.3cm} \sum\limits_{n=1}^{\infty} \Pr\big(\bigcup\limits_{i\in\mathcal{M}} \bigcup\limits_{k\in\mathcal{G}_i} \bigcup\limits_{l=n}^{\infty}(\widehat{D}_{i,k}(l)<D_{i,k} \mbox{\;or\;} \\
    \vspace{0.0cm} \hspace{3.9cm} \widehat{D}_{i,k}(l)> D_{i,k}^{(\max)}) \mbox{\;or\;} \\
    \vspace{0.0cm} \hspace{1.8cm} \bigcup\limits_{i\in\mathcal{M}} \bigcup\limits_{k\nin\mathcal{G}_i}\bigcup\limits_{l=n}^{\infty}(\widehat{D}_{i,k}(l)<D_{i,k})
    \big)
    \\\leq
    \vspace{0.0cm} \hspace{0.0cm} \sum\limits_{i\in\mathcal{M}}     \sum\limits_{k\in\mathcal{G}_i}\sum\limits_{n=1}^{\infty}        \sum\limits_{l=n}^{\infty}
    \Pr\big(\widehat{D}_{i,k}(l)<D_{i,k} \mbox{\;or\;}              \widehat{D}_{i,k}(l)> D_{i,k}^{(\max)}\big) \\
    \vspace{0.0cm} \hspace{0.3cm} +\sum\limits_{i\in\mathcal{M}}     \sum\limits_{k\nin\mathcal{G}_i}\sum\limits_{n=1}^{\infty}     \sum\limits_{l=n}^{\infty}
    \Pr\big(\widehat{D}_{i,k}(l)<D_{i,k}\big) $\\
\fi
Note that if we show that
\beq
\label{eq:Bound}
\bea{l}
 \Pr\big(\widehat{D}_{i,k}(l)<D_{i,k} \mbox{\;or\;} 
 \widehat{D}_{i,k}(l)> D_{i,k}^{(\max)}\big)\leq C\cdot l^{-(2+\delta)}
\ena
\eeq
\vspace{0.0cm} for some constants $C>0, \delta > 0$ for all $i\in\mathcal{M},k\in\mathcal{G}_i$ for all $l\geq n$, then we get: \vspace{0.0cm}\\
\ifOneCol
$ \displaystyle\sum\limits_{i\in\mathcal{M}} \sum\limits_{k\in\mathcal{G}_i}\sum\limits_{n=1}^{\infty} \sum\limits_{l=n}^{\infty}
\Pr \big(\widehat{D}_{i,k}(l)<D_{i,k} \mbox{\;or\;}  \widehat{D}_{i,k}(l)> D_{i,k}^{(\max)}\big)
\leq
M K C\left[\sum\limits_{l=1}^{\infty} l^{-(2+ \delta)}+\sum\limits_{n=2}^{\infty}\sum\limits_{l=n}^{\infty} l^{-(2+ \delta)}\right]
\\ \vspace{0.0cm} \hspace{-0.8cm} 
 \leq
  MK C\left[\sum\limits_{l=1}^{\infty} l^{-(2+ \delta)}+\sum\limits_{n=2}^{\infty}\int\limits_{n-1}^{\infty} l^{-(2+ \delta)}dl\right]=
MK C\left[\sum\limits_{l=1}^{\infty} l^{-(2+ \delta)}+\frac{1}{1+\delta}\sum\limits_{n=2}^{\infty}(n-1)^{-(1+\delta)}\right]<\infty$,\vspace{0.0cm}\\
\else
$ \displaystyle\sum\limits_{i\in\mathcal{M}} \sum\limits_{k\in\mathcal{G}_i}\sum\limits_{n=1}^{\infty} \sum\limits_{l=n}^{\infty}
\Pr \big(\widehat{D}_{i,k}(l)<D_{i,k} \mbox{\;or\;}  \widehat{D}_{i,k}(l)> D_{i,k}^{(\max)}\big)
\\\leq
\vspace{0.1cm} \hspace{0.3cm} M K \cdot C\left[\sum\limits_{l=1}^{\infty} l^{-(2+ \delta)}+\sum\limits_{n=2}^{\infty}\sum\limits_{l=n}^{\infty} l^{-(2+ \delta)}\right]
\\\leq
\vspace{0.1cm} \hspace{0.3cm} MK \cdot C\left[\sum\limits_{l=1}^{\infty} l^{-(2+ \delta)}+\sum\limits_{n=2}^{\infty}\int\limits_{n-1}^{\infty} l^{-(2+ \delta)}dl\right]
\\=
\vspace{0.1cm} \hspace{0.3cm} MK \cdot C\left[\sum\limits_{l=1}^{\infty} l^{-(2+ \delta)}+\frac{1}{1+\delta}\sum\limits_{n=2}^{\infty}(n-1)^{-(1+\delta)}\right]<\infty$,\vspace{0.2cm}\\
\fi
which is bounded independent of $t$. Similarly, showing that $\Pr\big(\widehat{D}_{i,k}(l)<D_{i,k}\big)\leq C\cdot l^{-(2+\delta)}$ for some constants $C, \delta > 0$ for all $i\in\mathcal{M},k\nin\mathcal{G}_i$ for all $j\geq n$ completes the statement.
\vspace{0.0cm}
We start bounding (\ref{eq:Bound}). We look at the first inequality of (\ref{eq:Bound}) for user $i$ with channel $k \in \mathcal{M}_i \cap \mathcal{T}_k$. 
The event  $\widehat{D}_{i,k}(t)<D_{i,k}$ implies: \vspace{0.0cm}\\
\ifOneCol
$\vspace{0.0cm} \hspace{0.0cm} \max \bigg\{ \Delta_{\min}^2, \min \big\{ \displaystyle \min_{\ell \neq k} \{ (\bar{s}_{i,k}(t)-\bar{s}_{i,\ell}(t))^2 \} -\epsilon, (\bar{s}_{i,k}(t) - \max_{j \neq i} \bar{s}_{j,k}(t))^2 -\epsilon \big\} \bigg\} \\
\vspace{0.0cm} \hspace{0.2cm} > \min \big\{ \displaystyle \min_{\ell \neq k} \{ (\mu_{i,k}-\mu_{i,\ell})^2 \}, ( \mu_{i,k} - \max_{j \neq i} \mu_{j,k})^2 \big\},\\ $
which after algebraic manipulations implies that at least one of the following holds:\vspace{0.0cm}
\begin{center}
$\vspace{0.0cm} \hspace{0.5cm} \displaystyle \min_{\ell \neq k} \{ (\bar{s}_{i,k}(t)-\bar{s}_{i,\ell}(t))^2 \} -\epsilon > \displaystyle \min_{\ell \neq k} \{ (\mu_{i,k}-\mu_{i,\ell})^2 \}$ \\
$\vspace{0.0cm} \hspace{0.5cm} (\bar{s}_{i,k}(t) - \max_{j \neq i} \bar{s}_{j,k}(t))^2 -\epsilon > ( \mu_{i,k} - \max_{j \neq i} \mu_{j,k})^2.
$
\end{center}
Similarly, the second inequality of (\ref{eq:Bound}) implies one of the following: \vspace{0.0cm}
\begin{center}
$\vspace{0.0cm} \hspace{0.3cm} \displaystyle \min_{\ell \neq k} \{ (\bar{s}_{i,k}(t)-\bar{s}_{i,\ell}(t))^2 \} -\epsilon < \displaystyle \min_{\ell \neq k} \{ (\mu_{i,k}-\mu_{i,\ell})^2 \} - 2\epsilon$ \\
$\vspace{0.0cm} \hspace{0.3cm} (\bar{s}_{i,k}(t) - \max_{j \neq i} \bar{s}_{j,k}(t))^2 -\epsilon < ( \mu_{i,k} - \max_{j \neq i} \mu_{j,k})^2 - 2\epsilon.$ 
\end{center}
Let $k^* = \displaystyle \text{arg} \min_{\ell \neq k} (\mu_{i,k}-\mu_{i,\ell} \})^2$ 
(i.e., $(\mu_{i,k}-\mu_{i,k^*} \})^2 = \displaystyle \min_{\ell \neq k} \{ (\mu_{i,k}-\mu_{i,\ell})^2 \}$).
Cascading the events written above we get \vspace{0.0cm}: \\
$\vspace{0.0cm} \hspace{0.3cm} \Pr \big(\widehat{D}_{i,k}(t)<D_{i,k} \mbox{\;or\;}  \widehat{D}_{i,k}(t)> D_{i,k}^{(\max)}\ \big) $
\begin{equation}
\vspace{0.0cm} \hspace{0.0cm} \leq \Pr \big( |(\bar{s}_{i,k}(t)-\bar{s}_{i,k^*}(t))^2 - (\mu_{i,k}-\mu_{i,k^*})^2| > \epsilon \big)  + \Pr \big( |(\bar{s}_{i,k}(t) - \max_{j \neq i} \bar{s}_{j,k}(t))^2- ( \mu_{i,k} - \max_{j \neq i} \mu_{j,k})^2| >\epsilon \big) \label{eq:1}.
\end{equation}
\else
$\vspace{0.3cm} \hspace{0.0cm} \max \bigg\{ \Delta_{\min}^2, \min \big\{ \displaystyle \min_{\ell \neq k} \{ (\bar{s}_{i,k}(t)-\bar{s}_{i,\ell}(t))^2 \} -\epsilon, \\
\vspace{0.3cm} \hspace{1.2cm} (\bar{s}_{i,k}(t) - \max_{j \neq i} \bar{s}_{j,k}(t))^2 -\epsilon \big\} \bigg\} \\
\vspace{0.3cm} \hspace{0.2cm} > \min \big\{ \displaystyle \min_{\ell \neq k} \{ (\mu_{i,k}-\mu_{i,\ell})^2 \}, ( \mu_{i,k} - \max_{j \neq i} \mu_{j,k})^2 \big\},\\ $
which after algebraic manipulations implies that at least one of the following holds:\vspace{0.3cm}\\
$\vspace{0.3cm} \hspace{0.5cm} \displaystyle \min_{\ell \neq k} \{ (\bar{s}_{i,k}(t)-\bar{s}_{i,\ell}(t))^2 \} -\epsilon > \displaystyle \min_{\ell \neq k} \{ (\mu_{i,k}-\mu_{i,\ell})^2 \} \\
\vspace{0.3cm} \hspace{0.5cm} (\bar{s}_{i,k}(t) - \max_{j \neq i} \bar{s}_{j,k}(t))^2 -\epsilon > ( \mu_{i,k} - \max_{j \neq i} \mu_{j,k})^2.
$ \\
Similarly, the second inequality of (\ref{eq:Bound}) implies one of the following: \vspace{0.3cm}\\
$\vspace{0.3cm} \hspace{0.3cm} \displaystyle \min_{\ell \neq k} \{ (\bar{s}_{i,k}(t)-\bar{s}_{i,\ell}(t))^2 \} -\epsilon < \displaystyle \min_{\ell \neq k} \{ (\mu_{i,k}-\mu_{i,\ell})^2 \} - 2\epsilon \\
\vspace{0.3cm} \hspace{0.3cm} (\bar{s}_{i,k}(t) - \max_{j \neq i} \bar{s}_{j,k}(t))^2 -\epsilon < ( \mu_{i,k} - \max_{j \neq i} \mu_{j,k})^2 - 2\epsilon.$ \\
Let $k^* = \displaystyle \text{arg} \min_{\ell \neq k} (\mu_{i,k}-\mu_{i,\ell} \})^2$ 
(i.e., $(\mu_{i,k}-\mu_{i,k^*} \})^2 = \displaystyle \min_{\ell \neq k} \{ (\mu_{i,k}-\mu_{i,\ell})^2 \}$).
Cascading the events written above we get \vspace{0.3cm}: \\
$\vspace{0.0cm} \hspace{0.3cm} \Pr \big(\widehat{D}_{i,k}(t)<D_{i,k} \mbox{\;or\;}  \widehat{D}_{i,k}(t)> D_{i,k}^{(\max)}\ \big) $
\begin{align}
\vspace{0.3cm} \hspace{0.0cm} \leq \Pr \big( |(\bar{s}_{i,k}(t)-\bar{s}_{i,k^*}(t))^2 - (\mu_{i,k}-\mu_{i,k^*})^2| > \epsilon \big) \nonumber \\
\vspace{0.3cm} \hspace{0.0cm} + \Pr \big( |(\bar{s}_{i,k}(t) - \max_{j \neq i} \bar{s}_{j,k}(t))^2- ( \mu_{i,k} - \max_{j \neq i} \mu_{j,k})^2| >\epsilon \big) \label{eq:1}.
\end{align}
\fi
Each of the terms in (\ref{eq:1}) is the probability of a deviation of the squared difference for two Markov chains' sample means from the squared difference of their expected means by an $\epsilon$. We look at the first term of (\ref{eq:1}).
Using conventional steps from set theory, it can be shown that: \vspace{0.0cm} \\
\ifOneCol
$\vspace{0.0cm} \hspace{0.0cm} \Pr \big( |(\bar{s}_{i,k}(t)-\bar{s}_{i,k^*}(t))^2 - (\mu_{i,k}-\mu_{i,k^*})^2| > \epsilon \big) \nonumber \\
\vspace{0.0cm} \hspace{0.0cm} \leq \big[ \Pr\big(|(\bar{s}_{i,k}(t)- \bar{s}_{i,k^*}(t)) [(\bar{s}_{i,k}(t)- \bar{s}_{i,k^*}(t)) -(\mu_{i,k}- \mu_{i,k^*})]|> \frac{\epsilon}{2} \big) \big] \\
\vspace{0.0cm} \hspace{0.0cm} + \big[ \Pr\big(|(\mu_{i,k}- \mu_{i,k^*})[(\bar{s}_{i,k}(t)- \bar{s}_{i,k^*}(t)) -(\mu_{i,k}- \mu_{i,k^*})]|> \frac{\epsilon}{2} \big) \big] \\ 
\vspace{0.0cm} \hspace{0.0cm} \leq \big[ \Pr\big(|(\bar{s}_{i,k}(t)- \bar{s}_{i,k^*}(t))-(\mu_{i,k}- \mu_{i,k^*})|>1 \big) +\Pr\big(|(\bar{s}_{i,k}(t)- \bar{s}_{i,k^*}(t))-(\mu_{i,k}- \mu_{i,k^*})|> \frac{\epsilon}{2(R+1)}\big) \\
\vspace{0.0cm} \hspace{-0.4cm} +\Pr\big(|(\mu_{i,k}- \mu_{i,k^*})+1|>R\big )\big] + \big[\vspace{0.0cm} \Pr \big(\mu_{i,k}>R' \big) +\Pr\big(|(\bar{s}_{i,k}(t)- \bar{s}_{i,k^*}(t))-(\mu_{i,k}- \mu_{i,k^*})|>\frac{\epsilon}{2(R'+1)} \big)\big] ,\\   
$ 
\else
$\vspace{0.3cm} \hspace{0.0cm} \Pr \big( |(\bar{s}_{i,k}(t)-\bar{s}_{i,k^*}(t))^2 - (\mu_{i,k}-\mu_{i,k^*})^2| > \epsilon \big) \nonumber \\
\vspace{0.2cm} \hspace{0.0cm} \leq \big[ \Pr\big(|(\bar{s}_{i,k}(t)- \bar{s}_{i,k^*}(t)) [(\bar{s}_{i,k}(t)- \bar{s}_{i,k^*}(t)) \\ 
\vspace{0.3cm} \hspace{4.7cm} -(\mu_{i,k}- \mu_{i,k^*})]|> \frac{\epsilon}{2} \big) \big] \\
\vspace{0.2cm} \hspace{0.0cm} + \big[ \Pr\big(|(\mu_{i,k}- \mu_{i,k^*})[(\bar{s}_{i,k}(t)- \bar{s}_{i,k^*}(t)) \\
\vspace{0.3cm} \hspace{4.7cm} -(\mu_{i,k}- \mu_{i,k^*})]|> \frac{\epsilon}{2} \big) \big] \\ 
\vspace{0.3cm} \hspace{0.0cm} \leq \big[ \Pr\big(|(\bar{s}_{i,k}(t)- \bar{s}_{i,k^*}(t))-(\mu_{i,k}- \mu_{i,k^*})|>1 \big)\\
\vspace{0.3cm} \hspace{0.0cm} +\Pr\big(|(\bar{s}_{i,k}(t)- \bar{s}_{i,k^*}(t))-(\mu_{i,k}- \mu_{i,k^*})|> \frac{\epsilon}{2(R+1)}\big) \\
\vspace{0.3cm} \hspace{0.0cm} +\Pr\big(|(\mu_{i,k}- \mu_{i,k^*})+1|>R\big )\big]\\
\vspace{0.3cm} \hspace{0.0cm} + \big[\vspace{0.3cm} \Pr \big(\mu_{i,k}>R' \big)\\
\vspace{0.3cm} +\Pr\big(|(\bar{s}_{i,k}(t)- \bar{s}_{i,k^*}(t))-(\mu_{i,k}- \mu_{i,k^*})|>\frac{\epsilon}{2(R'+1)} \big)\big] ,\\   
$ 
\fi
for every $R,R'>0$. We choose $R=R'=r_{\max}+1,$ hence the third and fourth terms are equal to $0$, and we get the concentration inequalities: \vspace{0.0cm} \\  
$\vspace{0.0cm} \hspace{0.0cm} \Pr \big( |(\bar{s}_{i,k}(t)-\bar{s}_{i,k^*}(t))^2 - (\mu_{i,k}-\mu_{i,k^*})^2| > \epsilon \big)$
\begin{align}
 < 6 \cdot \max \bigg\{
\Pr \big (|\bar{s}_{i,k}(t)-\mu_{i,k}|> \frac{\epsilon}{4(r_{\max}+2)} \big), \label{eq:2} \\
\Pr \big (|\bar{s}_{i,k^*}(t)-\mu_{i,k^*}|> \frac{\epsilon}{4(r_{\max}+2)} \big) \bigg\} \label{eq:3} .
\end{align}

Similar bounds can be obtained for the second term in (\ref{eq:1}).
To bound (\ref{eq:2}) and (\ref{eq:3}) we use Lezaud's results \cite{lezaud1998chernoff}: 

\begin{lemma}[\cite{lezaud1998chernoff}] Consider a finite-state, irreducible Markov chain $\{X_t\}_{t \geq 1}$ with state space $S$, matrix of transition probabilities $P$, an initial distribution $q$,  and stationary distribution $\pi$. Let
$N_\textbf{q}=\left \| (\frac{q^{(x)}}{\pi^{(x)}}, x \in S) \right \|_2 $.
Let $\widehat{P}=P'P $ be the multiplicative symmetrization of $P$ where $P'$ is the adjoint of $P$ on $l_2(\pi)$. Let $\epsilon= 1-\lambda_2$, where $\lambda_2$ is the second largest eigenvalue of the matrix $P'$. $\epsilon$ will be referred to as the eigenvalue gap of $P'$. Let $f:S \rightarrow \mathcal{R}$ be such that $\sum\limits_{y \in S} \pi_yf(y)=0, \quad \|f\|_2 \leq 1$ and $0 \leq \|f\|_2^2 \leq 1$ if $P'$ is irreducible. Then, for any positive integer $n$ and all $0<\lambda\leq 1$, we have:
$P \displaystyle \left(\frac{\sum\limits_{t=1} ^{n}f(X_t)}{n} \geq \lambda      \right) \leq N_\textbf{q}$ exp $[-\frac{n \lambda^2 \epsilon}{12}].$ \vspace{0.0cm}
\end{lemma}

Consider an initial distribution $\textbf{q}^{i,k}$ for channel $k$ of user $i$. We have:
\begin{center}
$ \displaystyle N_{\textbf{q}}^{(i,k)} = \left \| (\frac{q_{i,k}^x}{\pi_{i,k}^x}, x \in X^{i,k}) \right \|_2 \leq
  \sum\limits_{x \in X^{i,k}} \left \| \frac{q_{i,k}^x}{\pi_{i,k}^x} \right \|_2 \leq \frac{1}{\pi_{min}}. $
\end{center}
We point out that the sample rate mean $\bar{s}_{i,k}(t)$ is computed by $T^{(O)}_{i,k}(t)$ observation taken only from sub epochs DE in the exploration phases, thus the sample path that generated $\bar{s}_{i,k}(t)$ can be viewed as a sample path generated by a Markov chain with a transition matrix identical to the
original channel $\{i,k\}$, so we can apply Lezaud's result to bound (\ref{eq:2}) and (\ref{eq:3}). 
For equation (\ref{eq:2}): \\
we define $n_x^{i,k}(t)$ to be the number of occurrences of state $x$ on channel $k$ sensed by user $i$ up to time t. \vspace{0.0cm}\\
\ifOneCol
$\vspace{0.0cm} \Pr \big( \bar{s}_{i,k}(t) -\mu_{i,k} > \frac{\epsilon}{4(r_{\max}+2)} \big) =
 \Pr \big( \sum\limits_{x \in \mathcal{X}^{i,k}} x \cdot n_x^{i,k}(t)-T^{(O)}_{i,k}(t) \sum\limits_{x \in \mathcal{X}^{i,k}} x \cdot \pi_{i,k}^x > \frac{T^{(O)}_{i,k}(t)\cdot \epsilon}{4(r_{\max}+2)}\big) \\ =
\vspace{0.0cm} \hspace{0.0cm} \Pr \big(\sum\limits_{x \in \mathcal{X}^{i,k}} (x \cdot n_x^{i,k}(t)-T^{(O)}_{i,k}(t) x \cdot \pi_{i,k}^x) > \frac{T^{(O)}_{i,k}(t) \cdot \epsilon}{4(r_{\max}+2)}\big) \\ \leq
\vspace{0.0cm} \sum\limits_{x \in \mathcal{X}^{i,k}} \Pr \big(x \cdot n_x^{i,k}(t)-T^{(O)}_{i,k}(t) x \cdot \pi_{i,k}^x > \frac{T^{(O)}_{i,k}(t)\cdot \epsilon}{4(r_{\max}+2) |\mathcal{X}^{i,k}|} \big) \\ = 
\vspace{0.0cm} \sum\limits_{x \in \mathcal{X}^{i,k}} \Pr \big( n_x^{i,k}(t)-T^{(O)}_{i,k}(t) \cdot \pi_{i,k}^x > \frac{T^{(O)}_{i,k}(t)\cdot \epsilon}{4(r_{\max}+2) |\mathcal{X}^{i,k}| \cdot x} \big) \\ =
\vspace{0.0cm} \sum\limits_{x \in \mathcal{X}^{i,k}} \Pr \bigg( \frac{\sum \limits_{n=1}^t \textbf{1} (x_{i,k}(n)=x)-T^{(O)}_{i,k}(t) \pi_{i,k}^x } {\hat{\pi}_{i,k}^x \cdot T^{(O)}_{i,k}(t)} > \frac{T^{(O)}_{i,k}(t)\cdot \epsilon}{4(r_{\max}+2) |\mathcal{X}^{i,k}| \cdot x \hat{\pi}_{i,k}^x }\bigg) \\ \leq
\vspace{0.0cm} \hspace{0.0cm} |\mathcal{X}^{i,k}| \cdot N_\textbf{q}^{(i,k)}$ exp $\bigg(-T^{(O)}_{i,k}(t) \cdot \frac{\epsilon^2}{16(r_{\max}+2)^2 \cdot x^2 \cdot |\mathcal{X}^{i,k}|^2 \cdot (\hat{\pi}_{i,k}^x)^2} \cdot \frac{(1-\lambda_{i,k})}{12}\bigg),$ \\
\else
$\vspace{0.3cm} \Pr \big( \bar{s}_{i,k}(t) -\mu_{i,k} > \frac{\epsilon}{4(r_{\max}+2)} \big) \\ =
\vspace{0.3cm} \Pr \big( \sum\limits_{x \in \mathcal{X}^{i,k}} x \cdot n_x^{i,k}(t)-T^{(O)}_{i,k}(t) \sum\limits_{x \in \mathcal{X}^{i,k}} x \cdot \pi_{i,k}^x > \frac{T^{(O)}_{i,k}(t)\cdot \epsilon}{4(r_{\max}+2)}\big) \\ =
\vspace{0.3cm} \hspace{0.0cm} \Pr \big(\sum\limits_{x \in \mathcal{X}^{i,k}} (x \cdot n_x^{i,k}(t)-T^{(O)}_{i,k}(t) x \cdot \pi_{i,k}^x) > \frac{T^{(O)}_{i,k}(t) \cdot \epsilon}{4(r_{\max}+2)}\big) \\ \leq
\vspace{0.3cm} \sum\limits_{x \in \mathcal{X}^{i,k}} \Pr \big(x \cdot n_x^{i,k}(t)-T^{(O)}_{i,k}(t) x \cdot \pi_{i,k}^x > \frac{T^{(O)}_{i,k}(t)\cdot \epsilon}{4(r_{\max}+2) |\mathcal{X}^{i,k}|} \big) \\ = 
\vspace{0.3cm} \sum\limits_{x \in \mathcal{X}^{i,k}} \Pr \big( n_x^{i,k}(t)-T^{(O)}_{i,k}(t) \cdot \pi_{i,k}^x > \frac{T^{(O)}_{i,k}(t)\cdot \epsilon}{4(r_{\max}+2) |\mathcal{X}^{i,k}| \cdot x} \big) \\ =
\vspace{0.3cm} \sum\limits_{x \in \mathcal{X}^{i,k}} \Pr \bigg( \frac{\sum \limits_{n=1}^t \textbf{1} (x_{i,k}(n)=x)-T^{(O)}_{i,k}(t) \pi_{i,k}^x } {\hat{\pi}_{i,k}^x \cdot T^{(O)}_{i,k}(t)} \\  
\vspace{0.3cm} \hspace{4.5cm} > \frac{T^{(O)}_{i,k}(t)\cdot \epsilon}{4(r_{\max}+2) |\mathcal{X}^{i,k}| \cdot x \hat{\pi}_{i,k}^x }\bigg) \\ \leq
\vspace{0.3cm} \hspace{0.0cm} |\mathcal{X}^{i,k}| \cdot N_\textbf{q}^{(i,k)}$ exp $\bigg(-T^{(O)}_{i,k}(t) \cdot \frac{\epsilon^2}{16(r_{\max}+2)^2 \cdot x^2 \cdot |\mathcal{X}^{i,k}|^2 \cdot (\hat{\pi}_{i,k}^x)^2} \\ \vspace{0.3cm} \hspace{6.5cm} \cdot \frac{(1-\lambda_{i,k})}{12}\bigg),$ \\
\fi
and from (\ref{eq:Condition}), we have:
$T^{(O)}_{i,k}(t) > \frac{2}{I} \log(t)$ with $I$ defined in (\ref{eq:rate_function}). Thus,
\begin{align}
\displaystyle \Pr \big (|\bar{s}_{i,k}(t)-\mu_{i,k}|> \frac{\epsilon}{4(r_{\max}+2)} \big) \leq \frac{|X_{\max}|}{\pi_{\min}} \cdot t^{-2+\delta} \label{eq:4} .
\end{align}
The same bound can be obtained for (\ref{eq:3}), and with the same steps, for all terms in (\ref{eq:1}). The proof for all $i\in\mathcal{M},k\nin\mathcal{G}_i$ is similar, and thus Lemma \ref{lemma:T1} follows.
\hfill $\square$ \vspace{0.2cm} \\
We now bound the expected regret defined in (\ref{eq:regret}).
We divide the time horizon for $t<T_1$ and $t>T_1$. Since $T_1$ is finite (due to Lemma \ref{lemma:T1}), the regret for all $t<T_1$ results in a constant term $O(1)$ which is independent of $t$.
For $t>T_1$, we know that the adaptive exploration coefficient is no smaller than the deterministic exploration coefficient, and no larger than $D_{i,k}^{(\max)}$ defined in (\ref{eq:D_max}); i.e.,
\begin{equation}
 D_{i,k} \leq \widehat{D}_{i,k}(t) \leq D_{i,k}^{(\max)}, \label{eq:5}
\end{equation}
for all $i\in\mathcal{M}, k \in \mathcal{G}_i$ , and the LHS of the inequality for $i \in \mathcal{M}, k \in \mathcal{K}$. Thus, the exploration phases provides sufficient learning for the channel statistics (and the upper bound ensures that the channels are judiciously oversampled in the exploration phases). \\  
We continue bounding the regret for $t>T_1$:
\begin{align}
\displaystyle r(t) \leq (t-T_1) \cdot \sum \limits_{i=1}^M \mu_{i,S(i)} - \mathbb{E} [\sum\limits_{n=T_1+1}^{t} \sum\limits_{i=1}^M X_{i,a_i(n)}(n) ] \label{eq:6} .
\end{align}
For convenience, we will develop (\ref{eq:6}) between $n=1$ and $t$ with (\ref{eq:5}) (and the LHS for $k \nin \mathcal{G}_i$) holds for all $1 \leq n \leq t$, which upper bounds (\ref{eq:6}):\hspace{0.3cm}\\
$\vspace{0.0cm} \hspace{0.0cm} r(t) \leq (t-T_1) \cdot \sum \limits_{i=1}^M \mu_{i,S(i)} - \mathbb{E} [\sum\limits_{n=T_1+1}^{t} \sum\limits_{i=1}^M X_{i,a_i(n)}(n)]$ 
\begin{align}
\hspace{-2.3cm} \leq t \cdot \sum \limits_{i=1}^M \mu_{i,S(i)} - \mathbb{E} [\sum\limits_{n=1}^{t} \sum\limits_{i=1}^M X_{i,a_i(n)}(n)] \label{eq:7} .
\end{align}
We can rewrite (\ref{eq:7}) as: 
\begin{align}
\hspace{-1cm} r(t) & \leq \sum \limits_{i=1}^M \sum \limits_{k=1}^K \big( \mu_{i,k} \cdot E[T_{i,k}(t)] - E[\sum \limits_{n=1}^{t} X_{i,k}(n)] \big) \label{eq:transient}\\ 
& + \big( t \cdot \sum \limits_{i=1}^M \mu_{i,S(i)} - \sum \limits_{i=1}^M \sum \limits_{k=1}^K \mu_{i,k} \cdot E[T_{i,k}(t)] \big) \label{eq:suboptimal},
\end{align}
where $T_{i,k}(t)$ is the total number of transmission for user $i$ on channel $k$ up to time $t$ (and $X_{i,k}(n)=0$ if user $i$ did not try to access channel $k$ at time $n$). \\
Equation (\ref{eq:transient}) can be considered as the regret due to the transient effect (the initial state of the channel may not be given by the stationary distribution), and (\ref{eq:suboptimal}) is the regret caused by not playing the stable matching allocation. Both (\ref{eq:transient}) and (\ref{eq:suboptimal}) can be thought of as the sum of three different regret terms, corresponding to the three phases described in Section \ref{sec:DMASR}. We denote by $r^O(t), r^A(t) ,r^I(t)$ the regret caused in the exploration, allocation and exploitation  phases respectively; i.e., the regret can be written as:
\begin{align}
r(t) = r^O(t)+ r^A(t) + r^I(t) \label{eq:8}.
\end{align}
We next bound the regret in each of the three phases. \vspace{0.2cm}

\noindent\textbf{Regret in the exploration phases:}\vspace{0.1cm}\\
\vspace{0.1cm}
To bound the regret in the exploration phases, we first bound the number of exploration phases $n_O^{i,k}(t)$ for each user $i \in \mathcal{M}$ on each channel $k \in \mathcal{K}$ by time $t$. As described in Section (\ref{ssec:exploration}), the total number of samples from the exploration phases in sub epochs DE for user $i$ on channel $k$ up to time $t$ is:
\ifOneCol
$\displaystyle T_{i,k}^{(O)}(t) = \sum \limits_{n=1}^{n_O^{i,k}(t)} 4^{n-1} 
 = \frac{1}{3} (4^{n_O^{i,k}(t)}-1)$. \\
\else
\begin{center}
$\displaystyle T_{i,k}^{(O)}(t) = \sum \limits_{n=1}^{n_O^{i,k}(t)} 4^{n-1} 
 = \frac{1}{3} (4^{n_O^{i,k}(t)}-1)$. \\
\end{center}
\fi
Since we are in an exploration phase, from (\ref{eq:Condition}) together with (\ref{eq:5}), we have $T_{i,k}^{(O)}(t) < A_{i,k} \cdot \log(t)$ ($A_{i,k}$ is defined in (\ref{eq:20}).  
Hence, 
\beq
\label{eq:num_exploration}
\bea{l}
n_O^{i,k}(t) \leq \lfloor \log_4(3A_{i,k}\log(t)+1) \rfloor +1. 
\ena
\eeq  
We use the following lemma to show that the regret caused by channel switching is upper bounded by a constant independent of the number of transmissions on the channel in each phase.
\begin{lemma}[\cite{anantharam1987asymptotically}]
\label{lemma:ananthram}
Consider an irreducible, aperiodic Markov chain with state space $S$, a matrix of transition probabilities $P$, an initial distribution $\overrightarrow{q}$ which is positive in all states, and stationary distribution $\overrightarrow{\pi} (\pi_s$ is the stationary probability of state s). The state (reward) at time $t$ is denoted by $s(t)$. Let $\mu$ denote the mean reward. If we play the chain for an arbitrary time $T$, then there exists a value $A_p \leq (\min_{s \in S}\pi_s)^-1 \sum \limits_{s \in S} s$, such that: $E[\sum\limits_{t=1}^{T}s(t)-\mu T] \leq A_p$.
\end{lemma} 
Lemma \ref{lemma:ananthram} bounds the probability of a large deviation from the stationary distribution of a Markov chain (which we refer to as the transient effect). 
By the construction of the exploration phases described in Section (\ref{ssec:exploration}), in each exploration phase there is no channel switching (each channel has its own unique exploration phases), therefore (\ref{eq:transient}) in the exploration phases is bounded by: 
\beq
\label{eq:transient_regret_1}
\bea{l}
A_{\max} \cdot \big( \sum\limits_{i=1}^{M} \sum\limits_{k=1}^{K} (\lfloor \log_4(3A_{i,k}\log(t)+1) \rfloor +1) \big).
\ena
\eeq  
We next bound (\ref{eq:suboptimal}) in the exploration phases. Note that each user has its own exploration time, independent of the other users; i.e., when user $i$ explores, the other users (for which condition (\ref{eq:Condition}) holds) continue to exploit. However, user's $i$ exploration may affect other users exploring during that time due to collision.  
Specifically, when user $i$ explores channel $k$ it affects the regret in two ways. First, user $i$ does not transmit in its stable channel; hence, the regret is increased by $\mu_{i,S(i)}-\mu_{i,k}$. Second, if $k$ is a stable channel of another user, then because of the collision, the regret will increase by $\mu_{S^{-1}(k),k}$ ($S^{-1}(k)$ is the user for which channel $k$ is its stable channel
). Combining these two terms, we bound (\ref{eq:suboptimal}) in exploration phases by:
\beq
\label{eq:suboptimal_regret_1}
\bea{l}
\sum\limits_{i=1}^{M} \sum\limits_{k=1}^{K} \bigg(E[N_{i,k}^{(O)}(t)] \cdot (\mu_{i,S(i)} + \mu_{S^{-1}(k),k} - \mu_{i,k})  \bigg),
\ena
\eeq  
where $N_{i,k}^{(O)}(t)$ consists of the time indices from RE and DE, and depends on the mean hitting time of the channel due to the regenerative cycles. With (\ref{eq:num_exploration}) we have: \vspace{0.0cm} \\
\ifOneCol
$\vspace{0.0cm}E[N_{i,k}^{(O)}(t)] \leq \sum \limits_{n=0}^{n_O^{i,k}-1}(4^n+M^{i,k}_{\max}) = \frac{1}{3}(4^{n_O^{i,k}(t)}-1)+ M^{i,k}_{max} \cdot n_O^{i,k}(t)$ 
\beq
\label{eq:T_time_exploration}
\bea{l}
\vspace{0.0cm} \hspace{-2.0cm} \leq \frac{1}{3} [4(3A_{i,k}\cdot \log(t)+1)-1]  + M^{i,k}_{\max} \cdot \log_4(3A_{i,k}\log(t)+1).
\ena
\eeq 
\else
$\vspace{0.0cm}E[N_{i,k}^{(O)}(t)] \leq \sum \limits_{n=0}^{n_O^{i,k}-1}(4^n+M^{i,k}_{\max}) \\
\vspace{0.0cm} \hspace{0.0cm} = \frac{1}{3}(4^{n_O^{i,k}(t)}-1)+ M^{i,k}_{max} \cdot n_O^{i,k}(t)$ 
\beq
\label{eq:T_time_exploration}
\bea{l}
\vspace{0.0cm} \hspace{-2.0cm} \leq \frac{1}{3} [4(3A_{i,k}\cdot \log(t)+1)-1] \\
\vspace{0.0cm} \hspace{0.0cm} + M^{i,k}_{\max} \cdot \log_4(3A_{i,k}\log(t)+1).
\ena
\eeq 
\fi
Combining (\ref{eq:transient_regret_1}) and (\ref{eq:suboptimal_regret_1}) we can bound the first term in (\ref{eq:8}):
\beq
\label{eq:exploration_regret}
\bea{l}
\vspace{0.0cm} r^O(t) \leq A_{\max} \cdot \big( \sum\limits_{i=1}^{M} \sum\limits_{k=1}^{K} (\lfloor \log_4(3A_{i,k}\log(t)+1) \rfloor +1) \big) \\ + 
\vspace{0.0cm} \sum\limits_{i=1}^{M} \sum\limits_{k=1}^{K} \bigg(E[N_{i,k}^{(O)}(t)] \cdot (\mu_{i,S(i)} +  \mu_{S^{-1}(k),k} - \mu_{i,k})  \bigg),
\ena
\eeq 
which coincides with the first and second terms on the RHS of (\ref{eq:total_regret}).

\noindent\textbf{Regret in the allocation phases:}\vspace{0.1cm}\\
\vspace{0.1cm}
Since an allocation phase will only come after an exploration phase, the number of allocation phases by time $t$, $n_A(t)$ is bounded by the total number of exploration phases by time $t$; i.e., \begin{center}
$\vspace{0.0cm} n_A(t) \leq \sum \limits_{i=1}^M \sum \limits_{k=1}^K n_O^{i,k}(t),$  
\end{center}
and by using (\ref{eq:num_exploration}) we have: 
\beq
\label{eq:num_allocation}
\bea{l}
n_A(t) \leq \sum \limits_{i=1}^M \sum \limits_{k=1}^K \lfloor \log_4(3A_{i,k}\log(t)+1) \rfloor +1. 
\ena
\eeq  
Since the expected rates are unknown in our setting, the allocation phase is executed using the sample means.
To bound the expected time required for each allocation phase, we use proposition VI.4. in \cite{leshem2012multichannel}:

\begin{lemma}[\cite{leshem2012multichannel}]
\label{lemma:leshem}
Denote the expected delay to reach a stable matching configuration by $T_M$. There is some constant $C$ s.t. for every $M$ we have: \vspace{0.0cm}
\begin{center}
$T_M \leq C \log(M+1).$ \vspace{0.0cm}
\end{center}
\end{lemma}
Specifically, it was shown in \cite{leshem2012multichannel} that it is sufficient to choose $C=2e$ for the bound to hold. \\
Lemma \ref{lemma:leshem} states that each allocation phase is finite with respect to $t$, and only depends on the number of users. The total time in allocation phases by time $t$, denoted by $T_A(t)$, can be bounded by combining (\ref{eq:num_allocation}) with lemma \ref{lemma:leshem}: 
\ifOneCol
\beq
\label{eq:T_time_allocation}
\bea{l}
 E[T_A(t)] \leq \big(C \log(M+1)\big) \cdot \big( \sum \limits_{i=1}^M \sum \limits_{k=1}^K \lfloor \log_4(3A_{i,k}\log(t)+1) \rfloor +1 \big), 
\ena
\eeq   
\else
\beq
\label{eq:T_time_allocation}
\bea{l}
\vspace{0.3cm} \hspace{-1cm} E[T_A(t)] \leq \big(2C \log(M+1)\big) \\
\vspace{0.0cm} \hspace{0.3cm} \cdot \big( \sum \limits_{i=1}^M \sum \limits_{k=1}^K \lfloor \log_4(3A_{i,k}\log(t)+1) \rfloor +1 \big), 
\ena
\eeq   
\fi
with $\displaystyle C = 2e$. \vspace{0.0cm}

We now bound (\ref{eq:transient}) and (\ref{eq:suboptimal}) for the allocation phases.
In each allocation phase, the maximum number of channel switchings is $M \cdot M$; thus, the regret caused by the transient effect is bounded by:
\beq
\label{eq:transient_regret_2}
\bea{l}
A_{\max} \cdot M^2 \cdot \big( \sum\limits_{i=1}^{M} \sum\limits_{k=1}^{K} (\lfloor \log_4(3A_{i,k}\log(t)+1) \rfloor +1) \big).
\ena
\eeq  
and the regret due to sub-optimal allocation can be bounded by:    
\beq
\label{eq:suboptimal_regret_2}
\bea{l}
E[T_A(t)] \cdot \big( \sum\limits_{i=1}^{M} \mu_{i,S(i)} \big).
\ena
\eeq  
Combining (\ref{eq:transient_regret_2}), (\ref{eq:suboptimal_regret_2}) we have:
\ifOneCol
\beq
\label{eq:allocation_regret}
\bea{l}
\vspace{0.0cm} r^A(t) \leq A_{\max} \cdot M^2 \cdot \big( \sum\limits_{i=1}^{M} \sum\limits_{k=1}^{K} (\lfloor \log_4(3A_{i,k}\log(t)+1) \rfloor +1) \big) \\ 
\vspace{0.0cm} \hspace{0cm} + \big[ \big(C \log(M+1)\big)  \cdot \big( \sum \limits_{i=1}^M \sum \limits_{k=1}^K \lfloor \log_4(3A_{i,k}\log(t)+1) \rfloor +1 \big) \big] \cdot \big( \sum\limits_{i=1}^{M} \mu_{i,S(i)} \big),
\ena
\eeq 
\else
\beq
\label{eq:allocation_regret}
\bea{l}
\vspace{0.3cm} r^A(t) \leq A_{\max} \cdot M^2 \cdot \big( \sum\limits_{i=1}^{M} \sum\limits_{k=1}^{K} (\lfloor \log_4(3A_{i,k}\log(t)+1) \rfloor +1) \big) \\ 
\vspace{0.3cm} \hspace{0cm} + \big[ \big(C \log(M+1)\big)  \cdot \big( \sum \limits_{i=1}^M \sum \limits_{k=1}^K \lfloor \log_4(3A_{i,k}\log(t)+1) \rfloor +1 \big) \big] \\
\vspace{0.0cm} \hspace{2cm} \cdot \big( \sum\limits_{i=1}^{M} \mu_{i,S(i)} \big),
\ena
\eeq 
\fi
which coincides with the third and fourth terms in the RHS of (\ref{eq:total_regret}).

\noindent\textbf{Regret in the exploitation phases:}\vspace{0.1cm}\\
\vspace{0.0cm}
We first bound the number of exploitation phases up to time $t$.
As described in Section \ref{ssec:exploitation}, the number of time slots in the $n^{th}$ exploitation phase is $2 \cdot 4^{(n-1)}$. Thus we have:
\begin{center}
$\sum \limits_{n=1}^{n_I(t)} 2 \cdot 4^{n-1} = \frac{2}{3} (4^{n_I}-1) \leq t,$
\end{center}
which implies
\beq
\label{eq:num_exploitation}
\bea{l}
n_I \leq \lceil \log_4(\frac{3}{2}t+1) \rceil. 
\ena
\eeq 
During the exploitation phases, there are no channel switchings (each user exploits its stable channel). As a result, the regret caused by the transient effect in the exploitation phases is upper bounded by: 
\beq
\label{eq:transient_regret_3}
\bea{l}
A_{\max} \cdot \lceil \log_4(\frac{3}{2}t+1) \rceil.
\ena
\eeq  
It remains to bound the regret as a result of not playing the stable matching allocation (which we refer to as a sub-optimal allocation) in the exploitation phases. The event of playing a sub-optimal allocation in an exploitation phase occurs if the previous allocation phase results in a sub-optimal allocation, which occurs if one of the following takes place. The first is that user $i$ did not correctly identify the order of its $M$ best channels entering the allocation phase. This event would be denoted by $Y_i$. The second eventuality is when the user with the highest expected rate in channel $k$ was not identified correctly in the allocation phase. This event is denoted by $Z_k$.   
We write these events explicitly:
\begin{center}
$\vspace{0.0cm} \displaystyle Y_i(t_n)  = \bigcup\limits_{k \in \mathcal{M}_i} \bigcup\limits_{l \in \mathcal{K}} \big \{\bar{s}_{i,k}(t_n)< \bar{s}_{i,l}(t_n) | \mu_{i,k}>\mu_{i,l} \big \}$ \\ 
$\displaystyle Z_k(t_n)  = \bigcup\limits_{j \in \mathcal{T}_k}\big \{\bar{s}_{i,k}(t_n)<\bar{s}_{j,k}(t_n)|\mu_{i,k} = \max_{l \in \mathcal{T}_k} \mu_{l,k} \big \}, $
\end{center}
where $t_n$ denotes the starting time of the $n^{th}$ exploitation phase. Based on the above notations, the probability for a sub-optimal allocation ($P_S(n)$) in an exploitation phase at time $t_n$ is given by: 
\begin{center}
$\vspace{0.0cm} \hspace{0.3cm} \displaystyle P_S(n) \triangleq \Pr \big( \bigcup \limits_{i \in \mathcal{M}} Y_i(t_n) \mbox{\;or\;} \bigcup \limits_{k \in \mathcal{K}} Z_k(t_n) \big). $ 
\end{center}
The number of time slots in a sub-optimal allocation in the exploitation phases can be written as: \vspace{0.0cm} \ifOneCol
\begin{equation}
\displaystyle E[\tilde{T}(t)] = \sum \limits_{n=1}^{n_I(t)} 2 \cdot 4^{n-1} \cdot P_S(n) \leq  \sum \limits_{n=1}^{\lceil \log_4(\frac{3}{2}t+1) \rceil} 2 \cdot 4^{n-1} \cdot P_S(n) 
\displaystyle \leq \sum \limits_{n=1}^{\lceil \log_4(\frac{3}{2}t+1) \rceil} 3t_n \cdot P_S(n). \label{19}
\end{equation}
\else
$\\ \vspace{0.3cm} \hspace{0.3cm} \displaystyle E[\tilde{T}(t)] = \sum \limits_{n=1}^{n_I(t)} 2 \cdot 4^{n-1} \cdot P_S(n) \leq  \sum \limits_{n=1}^{\lceil \log_4(\frac{3}{2}t+1) \rceil} 2 \cdot 4^{n-1} \cdot P_S(n) $
\begin{align}
\vspace{0.3cm} \hspace{0.3cm}  \displaystyle \leq \sum \limits_{n=1}^{\lceil \log_4(\frac{3}{2}t+1) \rceil} 3t_n \cdot P_S(n). \label{19}
\end{align}
\fi
To complete Theorem \ref{th:regret}, we need to show that: 
\begin{align}
\displaystyle P_S(n) = \Pr \big( \bigcup \limits_{i \in \mathcal{M}} Y_i(t_n) \mbox{\;or\;} \bigcup \limits_{k \in \mathcal{K}} Z_k(t_n) \big) \leq B \cdot t_n^{-1} \label{eq:9}, 
\end{align}
for some $B>0$ (there is only a logarithmic number of terms in (\ref{19})). Using union bounds we have: 
\ifOneCol
\begin{equation}
\displaystyle \vspace{0.0cm} \hspace{0.0cm} \Pr \big( \bigcup \limits_{i \in \mathcal{M}} Y_i(t_n) \mbox{\;or\;} \bigcup \limits_{k \in \mathcal{K}} Z_k(t_n) \big) \leq  M^2 K \cdot \Pr \big( \bar{s}_{i,k}(t_n)< \bar{s}_{i,l}(t_n) | \mu_{i,k}>\mu_{i,l} \big) \label{eq:10} 
\end{equation}
\begin{equation}
+  MK \cdot \Pr \big( \bar{s}_{i,k}(t_n)<\bar{s}_{j,k}(t_n)|\mu_{i,k} = \max_{l \in \mathcal{T}_k} \mu_{l,k} \big) \label{eq:11}
\end{equation}
\else
$ \displaystyle \vspace{0.0cm} \hspace{0.0cm} \Pr \big( \bigcup \limits_{i \in \mathcal{M}} Y_i(t_n) \mbox{\;or\;} \bigcup \limits_{k \in \mathcal{K}} Z_k(t_n) \big) \\ $
\begin{align}
 \leq & M^2 K \cdot \Pr \big( \bar{s}_{i,k}(t_n)< \bar{s}_{i,l}(t_n) | \mu_{i,k}>\mu_{i,l} \big) \label{eq:10} \\
+ & MK \cdot \Pr \big( \bar{s}_{i,k}(t_n)<\bar{s}_{j,k}(t_n)|\mu_{i,k} = \max_{l \in \mathcal{T}_k} \mu_{l,k} \big) \label{eq:11}
\end{align}
\fi
To bound (\ref{eq:10}) and (\ref{eq:11}), we define $C_{t,v}= \sqrt{L \log(t)/v}$. Equation (\ref{eq:10}) implies that at least one of the following must hold \\
\ifOneCol
\noindent\begin{tabularx}{\textwidth}{@{}XXX@{}}
\begin{equation}
\bar{s}_{i,k}(t_n) \leq \mu_{i,k}- C_{t_n,T_{i,k}^{(O)}} \label{eq:12} 
\end{equation} &
\begin{equation}
 \bar{s}_{i,l}(t_n) \geq \mu_{i,l}+ C_{t_n,T_{i,l}^{(O)}} \label{eq:13}   
\end{equation} &
 \begin{equation}
    \mu_{i,k} <\mu_{i,l}+ C_{t_n,T_{i,l}^{(O)}}+ C_{t_n,T_{i,k}^{(O)}}. \label{eq:14}  
 \end{equation}
\end{tabularx}
\else
\begin{align}
\bar{s}_{i,k}(t_n) \leq \mu_{i,k}- C_{t_n,T_{i,k}^{(O)}} \label{eq:12} \\
\bar{s}_{i,l}(t_n) \geq \mu_{i,l}+ C_{t_n,T_{i,l}^{(O)}} \label{eq:13} \\
\mu_{i,k} <\mu_{i,l}+ C_{t_n,T_{i,l}^{(O)}}+ C_{t_n,T_{i,k}^{(O)}}. \label{eq:14} 
\end{align}
\fi
First we show that the probability for event (\ref{eq:14}) is zero. \vspace{0.0cm} \\
\ifOneCol
$\vspace{0.0cm} \hspace{0.4cm} \Pr \big(\mu_{i,k} <\mu_{i,l}+ C_{t_n,T_{i,l}^{(O)}}+ C_{t_n,T_{i,k}^{(O)}} \big) \displaystyle = \Pr \bigg( \mu_{i,k} - \mu_{i,l} < \sqrt{\frac{L \log t_n}{T_{i,l}^{(O)}(t_n)}}+\sqrt{\frac{L \log t_n}{T_{i,k}^{(O)}(t_n)}} \bigg) \\
\vspace{0.0cm} \hspace{0.0cm} \displaystyle \leq \Pr \bigg(\mu_{i,k} - \mu_{i,l} < 2\sqrt{\frac{L \log t_n}{\min\left\{T_{i,k}^{(O)}(t_n) , T_{i,l}^{(O)}(t_n)\right\}}}  \bigg) \\ 
\vspace{0.0cm} \hspace{-0.0cm} \displaystyle \leq \Pr \bigg( \min\left\{T_{i,k}^{(O)}(t_n) , T_{i,l}^{(O)}(t_n)\right\} < \frac{4L}{(\mu_{i,k} - \mu_{i,l})^2}\log(t_n) \bigg). \\
$
Combining (\ref{eq:5}) with (\ref{eq:Condition}) (which holds since we started an allocation phase), we have:
\begin{center}
$\vspace{0.0cm} \hspace{-0.2cm} \displaystyle T_{i,k}^{(O)}(t_n) > \frac{4L}{\displaystyle \min_{\ell \neq k} \{ (\mu_{i,k}-\mu_{i,\ell})^2\}}\log(t_n) \geq \frac{4L}{(\mu_{i,k} - \mu_{i,l})^2} \log(t_n)$ \\
$\vspace{0.0cm} \hspace{-0.2cm} \displaystyle T_{i,l}^{(O)}(t_n) > \frac{4L}{\displaystyle \min_{j \neq \ell} \{ (\mu_{i,l}-\mu_{i,j})^2 \}}\log(t_n) \geq \frac{4L}{(\mu_{i,k} - \mu_{i,l})^2} \log(t_n),$ 
\end{center}
\else
$\vspace{0.3cm} \hspace{0.4cm} \Pr \big(\mu_{i,k} <\mu_{i,l}+ C_{t_n,T_{i,l}^{(O)}}+ C_{t_n,T_{i,k}^{(O)}} \big) \\
\vspace{0.3cm} \hspace{0.0cm} \displaystyle = \Pr \bigg( \mu_{i,k} - \mu_{i,l} < \sqrt{\frac{L \log t_n}{T_{i,l}^{(O)}(t_n)}}+\sqrt{\frac{L \log t_n}{T_{i,k}^{(O)}(t_n)}} \bigg) \\
\vspace{0.3cm} \hspace{0.0cm} \displaystyle \leq \Pr \bigg(\mu_{i,k} - \mu_{i,l} < 2\sqrt{\frac{L \log t_n}{\min\left\{T_{i,k}^{(O)}(t_n) , T_{i,l}^{(O)}(t_n)\right\}}}  \bigg) \\ 
\vspace{0.3cm} \hspace{-0.3cm} \displaystyle \leq \Pr \bigg( \min\left\{T_{i,k}^{(O)}(t_n) , T_{i,l}^{(O)}(t_n)\right\} < \frac{4L}{(\mu_{i,k} - \mu_{i,l})^2}\log(t_n) \bigg). \\
$
Combining (\ref{eq:5}) with (\ref{eq:Condition}) (which holds since we started an allocation phase), we have: \vspace{0.3cm}\\
$\vspace{0.3cm} \hspace{-0.2cm} \displaystyle T_{i,k}^{(O)}(t_n) > \frac{4L}{\displaystyle \min_{\ell \neq k} \{ (\mu_{i,k}-\mu_{i,\ell})^2\}}\log(t_n) \\
\vspace{0.3cm} \hspace{4cm} \geq \frac{4L}{(\mu_{i,k} - \mu_{i,l})^2} \log(t_n) \\
\vspace{0.3cm} \hspace{-0.2cm} \displaystyle T_{i,l}^{(O)}(t_n) > \frac{4L}{\displaystyle \min_{j \neq \ell} \{ (\mu_{i,l}-\mu_{i,j})^2 \}}\log(t_n) \\ \vspace{0.3cm} \hspace{4cm} \geq \frac{4L}{(\mu_{i,k} - \mu_{i,l})^2} \log(t_n),$ \\
\fi
which ensures that the probability of (\ref{eq:14}) is zero. Note that here we used the fact that $D_{i,k} \geq D_{i,k}^{(R)}. $ \\
We now bound (\ref{eq:12}) and (\ref{eq:13}) using Lezaud's result (Lemma \ref{lemma:ananthram}). With similar steps as used above to bound (\ref{eq:2}), we can show:
\begin{align}
\vspace{0.0cm} \hspace{-0.2cm} &\Pr \big(\bar{s}_{i,k}(t_n) \leq \mu_{i,k}- C_{t_n,v_{i,k}} \big) 
\leq  \frac{|\mathcal{X}^{i,k}|}{\pi_{\min}} t^{- \frac{L \bar{\lambda}_{\min}}{28 X_{\max}^2 r_{\max}^2 \hat{\pi}_{\max}^2 } } \label{eq:15} \\
\vspace{0.0cm} \hspace{-0.2cm} &\Pr \big(\bar{s}_{i,l}(t_n) \geq \mu_{i,l}+ C_{t_n,v_{i,l}} \big) 
\leq  \frac{|\mathcal{X}^{i,l}|}{\pi_{\min}} t^{- \frac{L \bar{\lambda}_{\min}}{28 X_{\max}^2 r_{\max}^2 \hat{\pi}_{\max}^2 }}. \label{eq:16} 
\end{align}
Using (\ref{eq:L}), (\ref{eq:10}) is bounded by:
\ifOneCol
\begin{equation}
M^2 K \cdot \Pr \big( \bar{s}_{i,k}(t_n)< \bar{s}_{i,l}(t_n) | \mu_{i,k}>\mu_{i,l} \big)  \leq M^2 K \cdot \frac{2X_{\max}}{\pi_{\min}} \cdot t^{-1} \label{eq:17}.
\end{equation}
\else
\begin{align}
\vspace{0.0cm} \hspace{-0.3cm}&M^2 K \cdot \Pr \big( \bar{s}_{i,k}(t_n)< \bar{s}_{i,l}(t_n) | \mu_{i,k}>\mu_{i,l} \big) \nonumber \\
\vspace{0.0cm} \hspace{-0.3cm} \leq &M^2 K \cdot \frac{2X_{\max}}{\pi_{\min}} \cdot t^{-1} \label{eq:17}.
\end{align}
\fi
Equation (\ref{eq:11}) can be bounded using similar techniques, this time using the fact that $D_{i,k} \geq D_{i,k}^{(C)}$, and we can bound (\ref{eq:9}):
\begin{align}
\displaystyle \Pr \big( \bigcup \limits_{i \in \mathcal{M}} Y_i(t_n) \mbox{\;or\;} \bigcup \limits_{k \in \mathcal{K}} Z_k(t_n) \big)
\leq (M^2K+MK) \frac{2X_{\max}}{\pi_{\min}} \cdot t^{-1} \label{eq:18}.
\end{align}
With (\ref{eq:18}) we can bound (\ref{19}), and therefore the regret due to sub-optimal allocation in the exploitation phases is bounded by: 
\beq
\label{eq:suboptimal_regret_3}
\bea{l}
\displaystyle 3 \big(\sum\limits_{i=1}^{M} \mu_{i,S(i)}\big) (M^2K+MK) \frac{2X_{\max}}{\pi_{\min}} \cdot \lceil \log_4(\frac{3}{2}t+1) \rceil. 
\ena
\eeq  
By combining (\ref{eq:suboptimal_regret_3}) with (\ref{eq:transient_regret_3}), the total regret in the exploitation phases is:
\ifOneCol
\beq
\label{eq:exploitation_regret_3}
\bea{l}
\vspace{0.0cm} \hspace{0.0cm} \displaystyle r^I(t) \leq A_{\max} \cdot \lceil \log_4(\frac{3}{2}t+1) \rceil \displaystyle + 3 \big(\sum\limits_{i=1}^{M} \mu_{i,S(i)}\big) (M^2K+MK) \frac{2X_{\max}}{\pi_{\min}} \cdot \lceil \log_4(\frac{3}{2}t+1) \rceil,
\ena
\eeq   
\else
\beq
\label{eq:exploitation_regret_3}
\bea{l}
\vspace{0.0cm} \hspace{0.0cm} \displaystyle r^I(t) \leq A_{\max} \cdot \lceil \log_4(\frac{3}{2}t+1) \rceil \\
\vspace{0.0cm} \hspace{0.8cm} \displaystyle + 3 \big(\sum\limits_{i=1}^{M} \mu_{i,S(i)}\big) (M^2K+MK) \frac{2X_{\max}}{\pi_{\min}} \cdot \lceil \log_4(\frac{3}{2}t+1) \rceil,
\ena
\eeq   
\fi
which coincides with the two last terms on the RHS of (\ref{eq:total_regret}). \\  
\bibliographystyle{ieeetr}

\end{document}